\documentclass{aa}
\usepackage{rotating}
\usepackage{graphicx}
\usepackage{txfonts}
\begin{document}

\title{Spectroscopic and photometric variability of \newline the O9.5\,Vp star HD\,93521\thanks{Based on observations collected at the Observatoire de Haute Provence (France), the Flemish 1.2\,m Mercator telescope at the Roque de los Muchachos observatory (La Palma, Spain) and the Observatorio Astron\'omico Nacional of San Pedro M\'artir (Mexico).}}
\author{G.\ Rauw\inst{1}\fnmsep\thanks{Research Associate FRS-FNRS (Belgium)} \and M.\ De Becker\inst{1}\fnmsep\thanks{Postdoctoral Researcher FRS-FNRS (Belgium)} \and H.\ van Winckel\inst{2} \and C.\ Aerts\inst{2} \and P.\ Eenens\inst{3} \and\\ K.\ Lefever\inst{2} \and B.\ Vandenbussche\inst{2} \and N.\ Linder\inst{1} \and Y.\ Naz\'e\inst{1}\fnmsep$^{\star\star\star}$ 
\and E.\ Gosset\inst{1}\fnmsep$^{\star\star}$}
\offprints{G.\ Rauw}
\mail{rauw@astro.ulg.ac.be}
\institute{Institut d'Astrophysique et de G\'eophysique, Universit\'e de Li\`ege, All\'ee du 6 Ao\^ut, B\^at B5c, 4000 Li\`ege, Belgium \and
Instituut voor Sterrenkunde, K.U.\ Leuven, Celestijnenlaan 200D, 3001 Leuven, Belgium \and Departamento de Astronom\'{\i}a, Universidad de Guanajuato, Apartado 144, 36000 Guanajuato, GTO, Mexico}
\date{Received date / Accepted date}
\abstract{}{The line profile variability and photometric variability of the O9.5\,Vp star \object{HD\,93521} are examined in order to establish the properties of the non-radial pulsations in this star.}{Fourier techniques are used to characterize the modulations of the He\,{\sc i} $\lambda\lambda$\,5876, 6678 and H\,$\alpha$ lines in several spectroscopic time series and to search for variations in a photometric time series.}{Our spectroscopic data confirm the existence of two periods of 1.75 and 2.89\,hr. The line profiles, especially those affected by emission wings, exhibit also modulations on longer time scales, but these are epoch-dependent and change from line to line. Unlike previous claims, we find no unambiguous signature of the rotational period in our data, nor of a third pulsation period (corresponding to a frequency of 2.66\,d$^{-1}$).}{HD\,93521 very likely exhibits non-radial pulsations with periods of 1.75 and 2.89\,hr with $l \simeq 8 \pm 1$ and  $l \simeq 4 \pm 1$ respectively. No significant signal is found in the first harmonics of these two periods. The 2.89\,hr mode is seen at all epochs and in all lines investigated, while the visibility of the 1.75\,hr mode is clearly epoch dependent. Whilst light variations are detected, their connection to these periodicities is not straightforward.}
\authorrunning{G.\ Rauw et al.\ }
\maketitle

\section{Introduction \label{intro}}
Time resolved spectroscopy of O-type stars has shown that absorption line profile variability at the level of a few per cent is a common feature (e.g.\ Fullerton et al.\ \cite{FGB}). Various mechanisms, including magnetic fields, stochastic structures at the base of the wind and non-radial pulsations have been proposed to explain this variability. Despite their rather subtle signature, the diagnostic potential of these phenomena is considerable.  Especially in the case of non-radial pulsations, the emerging discipline of asteroseismology offers the possibility to gain insight into the interiors of early-type stars. However, to characterize the nature of the phenomenon requires a rather long and well sampled time series of spectra with a high resolution and a high S/N (see e.g.\ Aerts et al.\,\cite{nuEri} for the case of the $\beta$\,Cephei variable $\nu$\,Eri). Up to now, such detailed studies have therefore been restricted to a few, rather bright and well-known O-stars such as $\zeta$\,Pup (Baade et al.\ \cite{Baade}) and $\zeta$\,Oph (Kambe et al.\ \cite{Kambe}).\\ 

In this context, the high Galactic latitude O-star HD\,93521 ($l_{\rm II} = 183.14^{\circ}$, $b_{\rm II} = 62.15^{\circ}$) is an extremely interesting target. Based on optical spectroscopy, Fullerton et al.\ (\cite{FGB2}) and Howarth \& Reid (\cite{HR}) found HD\,93521 to exhibit bumps at the 1\% level moving from blue to red across the profiles of the He\,{\sc i} lines, whilst no variability was detected in the He\,{\sc ii} lines. Fullerton et al.\ (\cite{FGB2}) as well as Howarth \& Reid (\cite{HR}) accordingly interpreted these features as the signature of non-radial pulsations with a period of 1.8\,hours. The He\,{\sc i} -- He\,{\sc ii} dichotomy was interpreted as arising from the substantial gravity darkening that favours He\,{\sc i} line formation near the cooler equatorial regions where the pulsational amplitude attains a maximum (Townsend \cite{Townsend}). The existence of non-radial pulsations was subsequently confirmed by Howarth et al.\ (\cite{HTC}) using {\it IUE} spectra and three different periods were identified. Recently, Rzaev \& Panchuk (\cite{RP}) reported on the existence of slightly different variability patterns between the strong and weak He\,{\sc i} lines. However, since the Rzaev \& Panchuk (\cite{RP}) data set covered only 2.7\,hours, it yields hardly any constraint on the properties of the pulsations.\\ 

HD\,93521 has one of the largest rotational velocities known among O-stars (341\,km\,s$^{-1}$, Penny \cite{Penny}; 432\,km\,s$^{-1}$, Howarth et al.\ \cite{HSHP}; 390\,km\,s$^{-1}$, see below in this paper). The stellar wind has an apparently low terminal velocity and is likely heavily distorted by rotation (e.g.\ Bjorkman et al.\ \cite{Bjorkman}). In the optical, the wind produces emission features in the wings of the H$\alpha$ line, although this line has so far never been reported as a pure emission feature. 
\begin{table*}
\caption{Summary of our observing runs (see text for the meaning of the different columns). The green and red wavelength ranges stand respectively for 5500 -- 5920 and 6530 -- 6710\,\AA, whilst SPM stands for the echelle spectra taken at San Pedro M\'artir. The last column yields the mean signal-to-noise ratio of the spectra gathered during the corresponding campaign. The last line refers to the photometric monitoring campaign that took place in coordination with the April 2005 spectroscopic campaign.\label{journal}}
\begin{center}
\begin{tabular}{l c c c c c c c}
\hline
\vspace*{-3mm}\\
Epoch & Domain & $\Delta{\rm T}$ (days) & N & $\overline{\Delta t}$ (days) & $\Delta\,\nu_{\rm nat}$ (d$^{-1})$ & $\nu_{\rm max}$ (d$^{-1})$ & $\overline{\rm S/N}$\\
\hline
February 1997 & red & 4.155 & 35 & $2.49 \times 10^{-2}$ & 0.241 & 20.1 & 165\\
April 2005 & red & 3.809 & 55 & $1.72 \times 10^{-2}$ & 0.263 & 29.1 & 280 \\
April 2005 & SPM & 5.096 & 90 & $0.84 \times 10^{-2}$ & 0.196 & 59.6 & 250 \\
February 2006 & green & 6.173 & 75 & $1.60 \times 10^{-2}$ & 0.162 & 31.3 & 650 \\
April 2006 & green & 0.180 & 13 & $1.49 \times 10^{-2}$ & 5.55 & 33.5 & 580 \\
April 2007 & green & 6.191 & 56 & $1.46 \times 10^{-2}$ & 0.162& 34.2 & 550 \\
\hline
April 2005 & $U\,B\,B_1\,B_2\,V\,V_1\,G$ & 28.221 & 378 & $0.31 \times 10^{-2}$ & $0.035$ & 161.3 & \\
\hline
\end{tabular}
\end{center}
\end{table*}

\begin{figure*}[t!]
\begin{minipage}{6.0cm}
\resizebox{6.0cm}{!}{\includegraphics{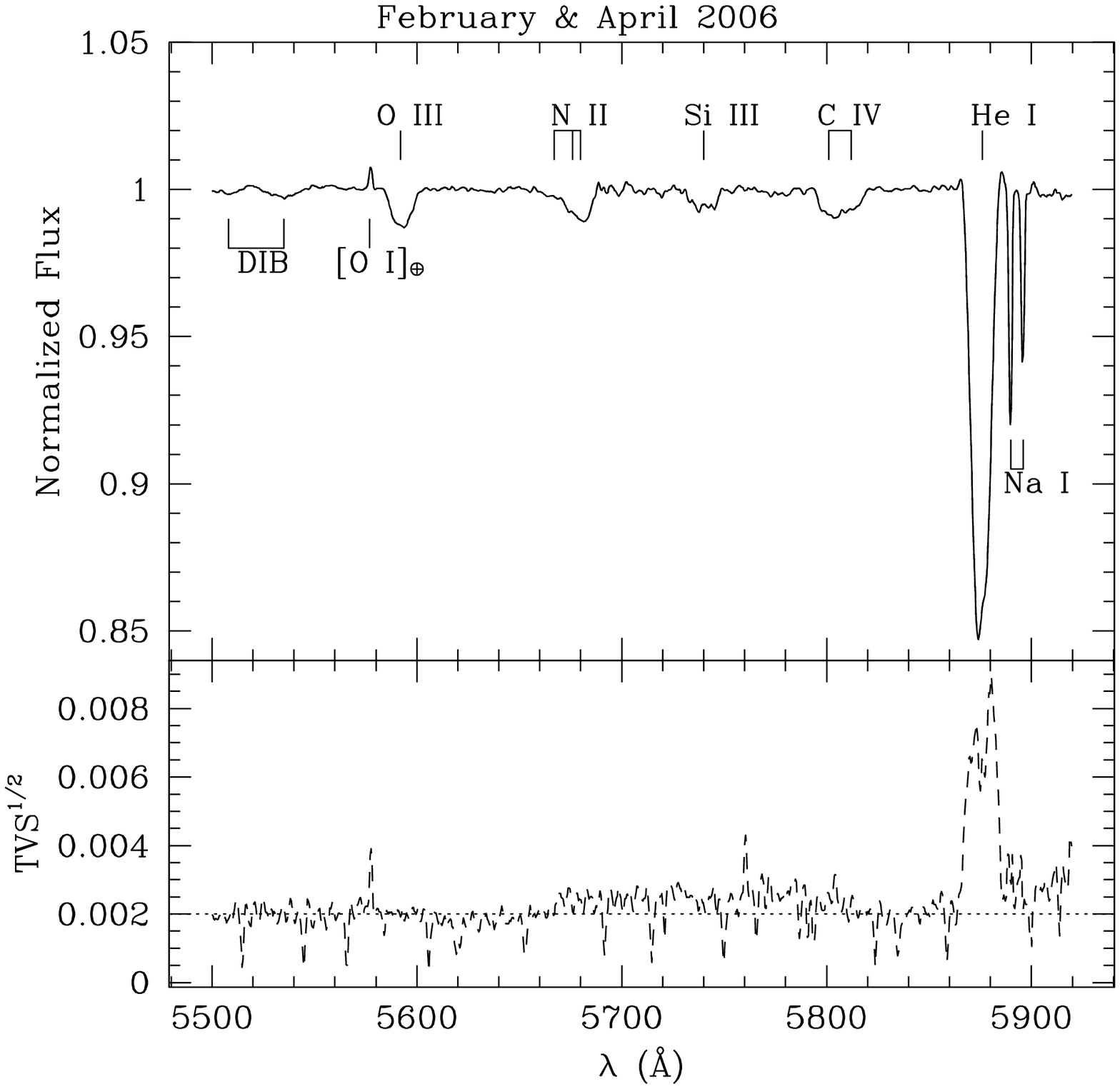}}
\end{minipage}
\begin{minipage}{6.0cm}
\resizebox{6.0cm}{!}{\includegraphics{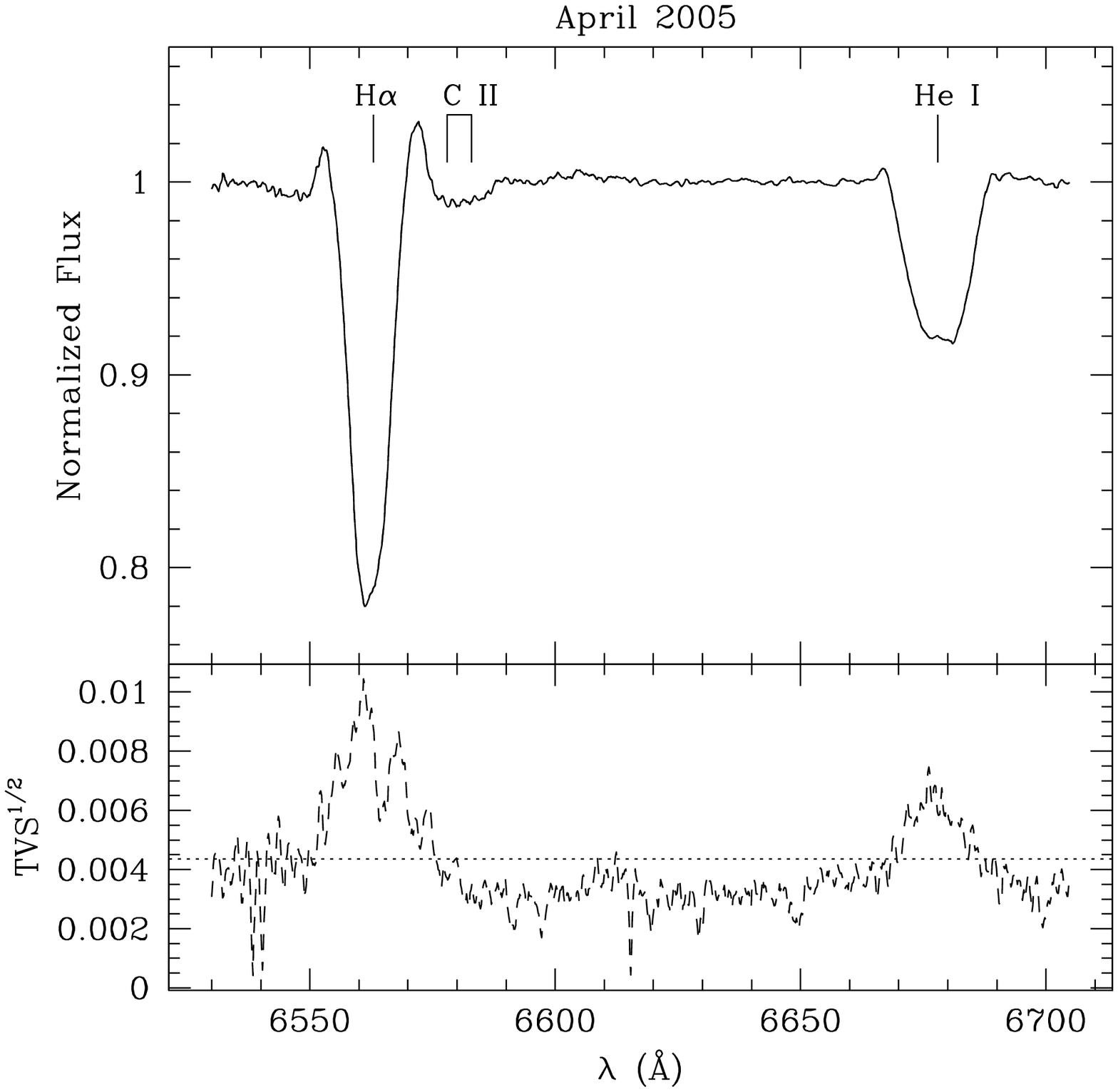}}
\end{minipage}
\begin{minipage}{6.0cm}
\resizebox{6.0cm}{!}{\includegraphics{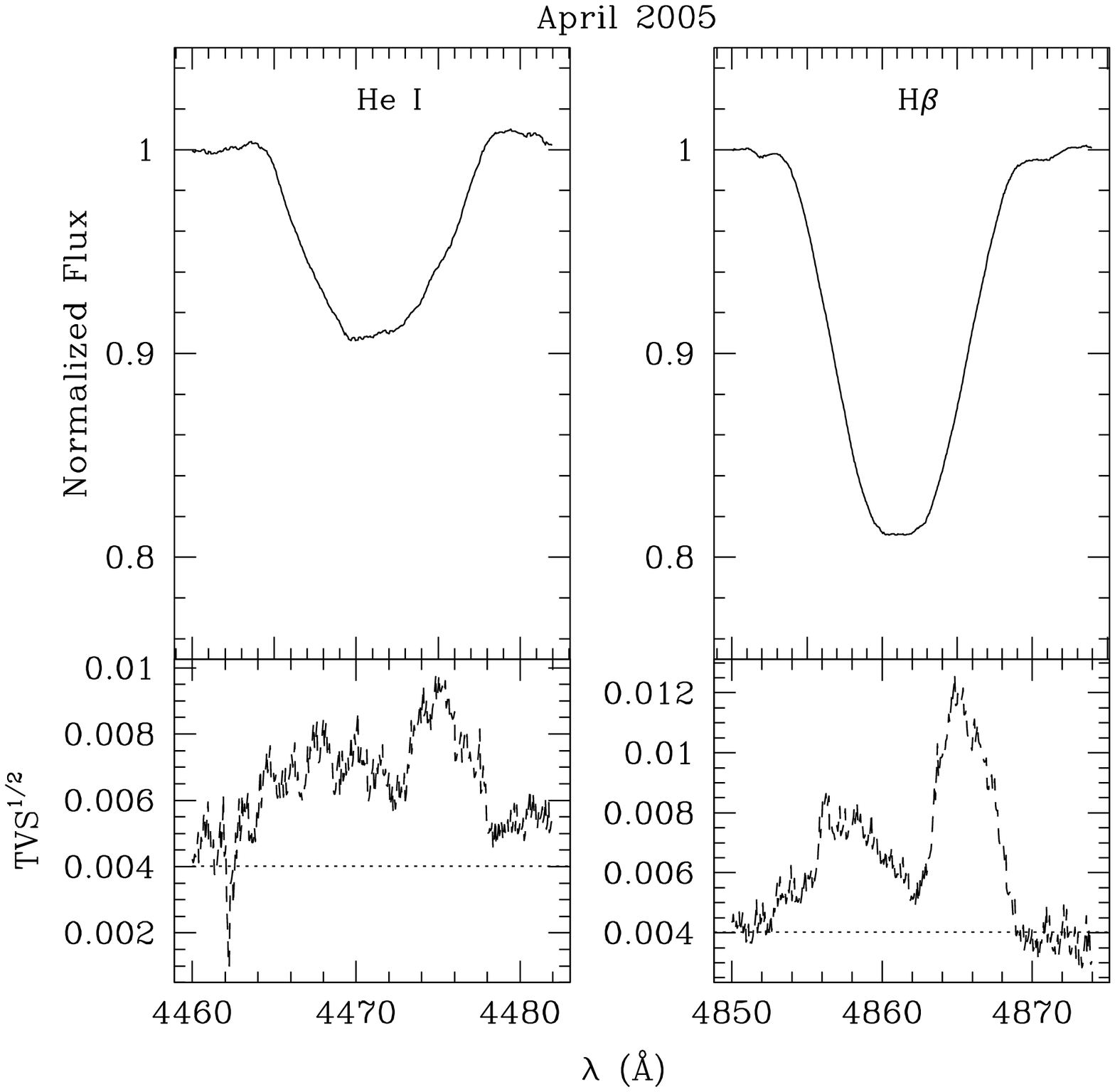}}
\end{minipage}
\caption{Average spectrum (top) and temporal variance spectrum (TVS, bottom) computed from the green, red and SPM spectra taken in February + April 2006 (left panel), April 2005 (middle) and April 2005 (right) respectively. The dotted lines yield the 99\% significance level for the variability evaluated following the approach of Fullerton et al.\ (\cite{FGB}).\label{average}}
\end{figure*}

While the optical spectrum of HD\,93521 leads to an O9.5\,V classification, the nature of this star has been subject to debate over many years. In fact, assuming a typical absolute magnitude for a Population I O9.5\,V star, HD\,93521 is located about 1.4\,kpc above the Galactic plane (Irvine \cite{Irvine}), far away from any site of recent massive star formation. While there is still some uncertainty concerning the motion of the star (towards or away from the Galactic plane, see Gies \cite{Gies}), it seems unlikely that HD\,93521 could have formed in the plane and subsequently moved to its current position. Furthermore, based on the rather low stellar wind terminal velocity of HD\,93521 and assuming that $v_{\infty} = 3 \times v_{esc}$, Ebbets \& Savage (\cite{ES}) concluded that this star was likely a low-mass evolved Population II object. However, Irvine (\cite{Irvine}) showed that the relation between $v_{\infty}$ and $v_{esc}$ does not hold for late O-type stars such as HD\,93521 and he proposed that the star is in fact a normal main sequence star that has formed in the Galactic halo (another example of a massive star born in the halo can be found in Heber et al.\ \cite{Heber}). Lennon et al.\ (\cite{Lennon}) measured the equivalent widths (EWs) of several metal lines in the spectrum of HD\,93521. Though these lines are washed out by rotational broadening and the EWs are affected by large uncertainties, their strengths are inconsistent with Population II metal abundances. Finally, Massa (\cite{Massa}) showed that the other peculiar features of this star (its unusually low UV continuum, its abnormally strong wind lines in the UV, its low excitation photospheric lines) can all be accounted for by the effect of gravity darkening in a `normal' Population I star rotating at 90\% of its breakup velocity and seen nearly equator on.\\

In this paper, we present the results of a spectroscopic and photometric monitoring campaign of HD\,93521. The aim of this project was to check the long-term stability of the periodicities identified in previous investigations. 

\section{Observations}
\subsection{Spectroscopy}
We observed HD\,93521 during five observing campaigns (in February 1997, April 2005, February and April 2006 as well as April 2007) with the Aur\'elie spectrograph at the 1.52\,m telescope of the Observatoire de Haute Provence (OHP). The 1997 and 2005 data were taken with a 1200 lines/mm grating blazed at 5000\,\AA\ and covered the wavelength domain from 6530 to 6710\,\AA\ with a resolving power of $R \simeq 21000$. The 2006 and 2007 data were obtained with a 600 lines/mm grating also blazed at 5000\,\AA\ and covering this time the region between 5500 and 5920\,\AA\ with $R \simeq 10000$. In 1997, the detector was a Thomson TH7832 linear array with 2048 pixels of 13\,$\mu$m each, whilst later observations used an EEV 42-20 CCD with $2048 \times 1024$ pixels. All the OHP data were reduced in the standard way (see Rauw et al.\ \cite{R03}) using the MIDAS software provided by ESO. To correct the observed spectra to first order for telluric (mainly water vapor) absorption lines, a template of the telluric spectrum was built from observations of a bright star at very different airmasses (usually around 1.2 and 3.0). The stars used for this purpose were $\nu$\,Ori (B3\,V) and HD\,100889 (B9.5\,Vn) in February 1997, HD\,177724 (A0\,Vn) in April 2005 and Regulus ($\alpha$\,Leo, B7\,V) in February 2006. Finally, the spectra were normalized using properly chosen continuum windows. Special care was taken to ensure that all the data were normalized in a homogeneous way to allow a self-consistent search for variability. 

In April 2005, we gathered a series of 90 spectra with the echelle spectrograph mounted on the 2.1\,m telescope at the Observatorio Astron\'omico Nacional of San Pedro M\'artir (SPM) in Mexico. The instrument covers the spectral domain between about 3800 and 6800\,\AA\ with a resolving power of 18000 at 5000\,\AA. The detector was a SITe CCD with $1024 \times 1024$ pixels of 24\,$\mu$m$^2$. The data were reduced using the echelle context of the MIDAS software and specific orders covering several important lines (He\,{\sc i} $\lambda\lambda$\,4471, 5876, H$\beta$ and H$\alpha$) were normalized using a set of selected continuum windows. Unfortunately, the He\,{\sc i} $\lambda$\,6678 line could not be studied because it fell too close to the edge of an order. 

A summary of the main characteristics of our various data sets is provided in Table\,\ref{journal}. For each run, $\Delta{\rm T}$ gives the time elapsed between our first and our last observation, while N is the total number of observations. $\overline{\Delta t}$ is the average time interval between two consecutive exposures during the same night. In light of the Fourier analysis in Section \ref{sect: fourier}, we further list the typical FWHM of a peak in the power spectrum $\Delta\,\nu_{\rm nat} = 1/\Delta{\rm T}$, as well as $\nu_{\rm max} = 1/(2\,\overline{\Delta t})$ which provides a rough indication of the highest frequencies that could be sampled with our time series.

\begin{figure}[thb!]
\resizebox{9.0cm}{!}{\includegraphics{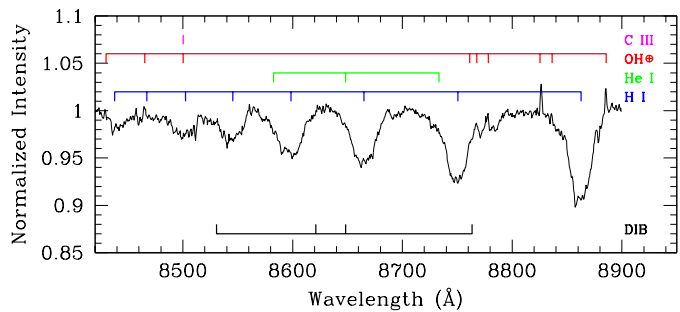}}
\caption{Mean spectrum of HD\,93521 over the 8420 -- 8900\,\AA\ wavelength domain. The ions responsible for the lines (either stellar, or telluric in the case of the OH emissions) are indicated above the spectrum, whilst the diffuse interstellar bands (DIBs) are indicated below.\label{GAIA}}
\end{figure}

During the April 2006 campaign, HD\,93521 was also observed in the near-infrared  with the Aur\'elie spectrograph. The data were obtained with a 300 lines/mm grating blazed at 6000\,\AA\ and covering the region between 8055 and 8965\,\AA\ with $R \simeq 6500$. Whilst the individual spectra in this wavelength domain do not have a sufficient quality to perform a meaningful variability analysis, we present the mean spectrum over the useful band from 8420 to 8900\,\AA\ in Fig.\,\ref{GAIA}. The spectrum is largely dominated by broad and shallow hydrogen absorption lines of the Paschen series (from Pa11 to at least Pa16) with some modest contributions from He\,{\sc i} and possibly He\,{\sc ii}, C\,{\sc iii} and N\,{\sc iii} lines. The RVS instrument onboard ESA's forthcoming astrometry mission GAIA will cover a sub-domain (from 8470 -- 8740\,\AA) of the band shown in Fig.\,\ref{GAIA} with a resolving power twice as large as that of our data to measure among other things the radial and rotational velocities. Our spectrum indicates that the only prominent features in the RVS spectra of rapidly rotating late O-type dwarfs are the Paschen lines.

\subsection{Photometry}
Time resolved photometry of HD\,93521 through the $(U\,B\,B_1\,B_2\,V\,V_1\,G)_G$ Geneva filter system (see e.g.\ Rufener \cite{Rufener}, Bessell \cite{Bessell}) was obtained with the Mercator telescope in April 2005. Mercator is a 1.2\,m semi-robotic telescope, situated at Roque de los Muchachos (La Palma Island, Spain). The telescope is equipped with the P\,7 photometer, which is a two channel photometer for quasi simultaneous 7 band measurements. The first channel (A) is centered on the star while the second channel (B) is centered on the sky. The position angle of the sky can be changed by turning the derotator while changing the distance between both channels needs manual interaction. The filter wheel turns at 4\,Hz and a chopper directs both channels alternatively to the photomultiplier. As such, the photomultiplier measures both beams A and B through the seven filters four times each second.

The strategy for performing the observations is oriented towards obtaining
high-precision photometry. In order to achieve this, we measure stars within a
range of 0.1 in air-mass $F_z$ for nights of good atmospheric conditions. It is
advantageous to measure a variable star with a period of the order of hours
frequently each night, so we have opted for the interval $F_z\,\in\,[1.05;1.15]$
to measure HD\,93521.  For the reduction process of such a type of observing
night, typically 2--3 standard stars of different colour are observed
each hour.
The determination of the extinction coefficients was done according to the
method outlined in Burki et al.\ (\cite{Burki}) and results in measurements with a typical accuracy of a few mmag for the colours of stars with visual magnitude
between 5 and 10. A total of 378 data points were gathered between 29 March and 27 April (UT dates). In our analysis we have restricted ourselves to those 315 data points that have weights of 3 or 4 (for good or exceptional quality measurements respectively, see Rufener \cite{Rufener}). Nights with these quality weights are characterized by standard deviations on the measurements of standard stars of less than 0.005\,mag.  

\begin{figure}[thb!]
\resizebox{9.0cm}{!}{\includegraphics{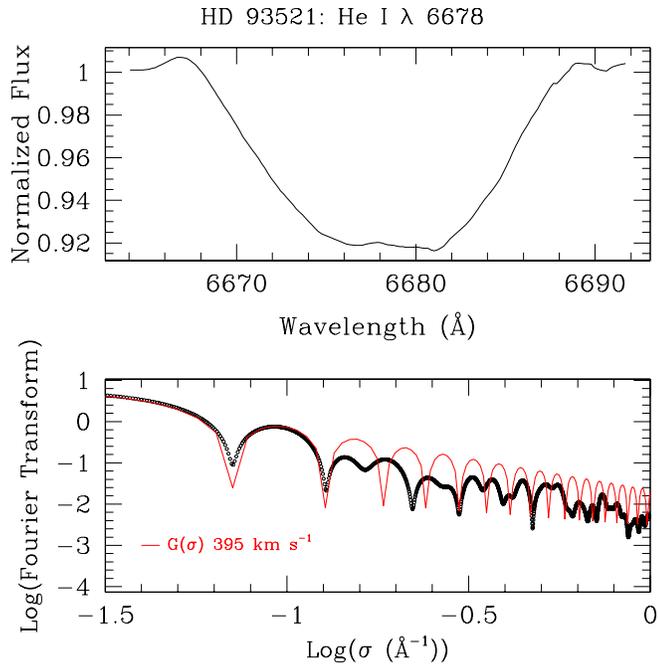}}
\caption{Determination of the rotational velocity of HD\,93521 using the Fourier technique. The top panel illustrates the mean profile of the He\,{\sc i} $\lambda$\,6678 line as observed in April 2005. The lower panel illustrates the resulting Fourier transform along with the Fourier transform $G(\sigma)$ of a rotational broadening function (see Gray \cite{Gray}) evaluated for $v\,\sin{i} = 395$\,km\,s$^{-1}$.\label{vsini}}
\end{figure}
\section{Results}
\subsection{Spectroscopy \label{spec}}
The average spectra of the star over the wavelength domains investigated here are shown in Fig.\,\ref{average}. Besides a few diffuse interstellar bands (DIBs), identified following the catalogue of interstellar features from Herbig (\cite{Herbig}), we note the broad absorption lines as well as the emission components of H$\alpha$ and He\,{\sc i} $\lambda$\,5876. Note that the He\,{\sc i} lines appear relatively strong compared to the H$\alpha$ line. This can be understood in view of the helium overabundance ($y = 0.18 \pm 0.03$) reported by Lennon et al.\ (\cite{Lennon}) and Howarth \& Smith (\cite{HS}).

In addition to the O\,{\sc iii} $\lambda$\,5592, C\,{\sc iv} $\lambda\lambda$\,5801, 5812 and He\,{\sc i} $\lambda$\,5876 absorptions that are typical for an O-type star, the spectrum in the region between 5500 and 5900\,\AA\ also reveals several unusual features, such as the N\,{\sc ii} $\lambda\lambda$\,5667, 5676, 5680 blend and the Si\,{\sc iii} $\lambda$\,5740 line. These lines are strong in the spectra of early B supergiants, but are usually absent from O-star spectra (see Walborn \cite{Walborn}). The same remark applies to the feature near the red emission wing of H$\alpha$ that we identify as C\,{\sc ii} $\lambda\lambda$\,6578, 6583 which are clearly seen in the spectra of early B-type stars. These particularities are reminiscent of the description of the {\it IUE} spectrum of the star given by Massa (\cite{Massa}) and thus likely reflect the strong temperature gradient that exists at the stellar surface as a result of gravity darkening. According to Massa (\cite{Massa}), the polar region has a temperature close to that of an O7 star, whilst the equatorial region is much cooler and experiences a lower surface gravity, hence producing a spectrum similar to that of an early B supergiant.

\begin{figure*}[htb!]
\begin{minipage}{9.0cm}
\resizebox{9.0cm}{!}{\includegraphics{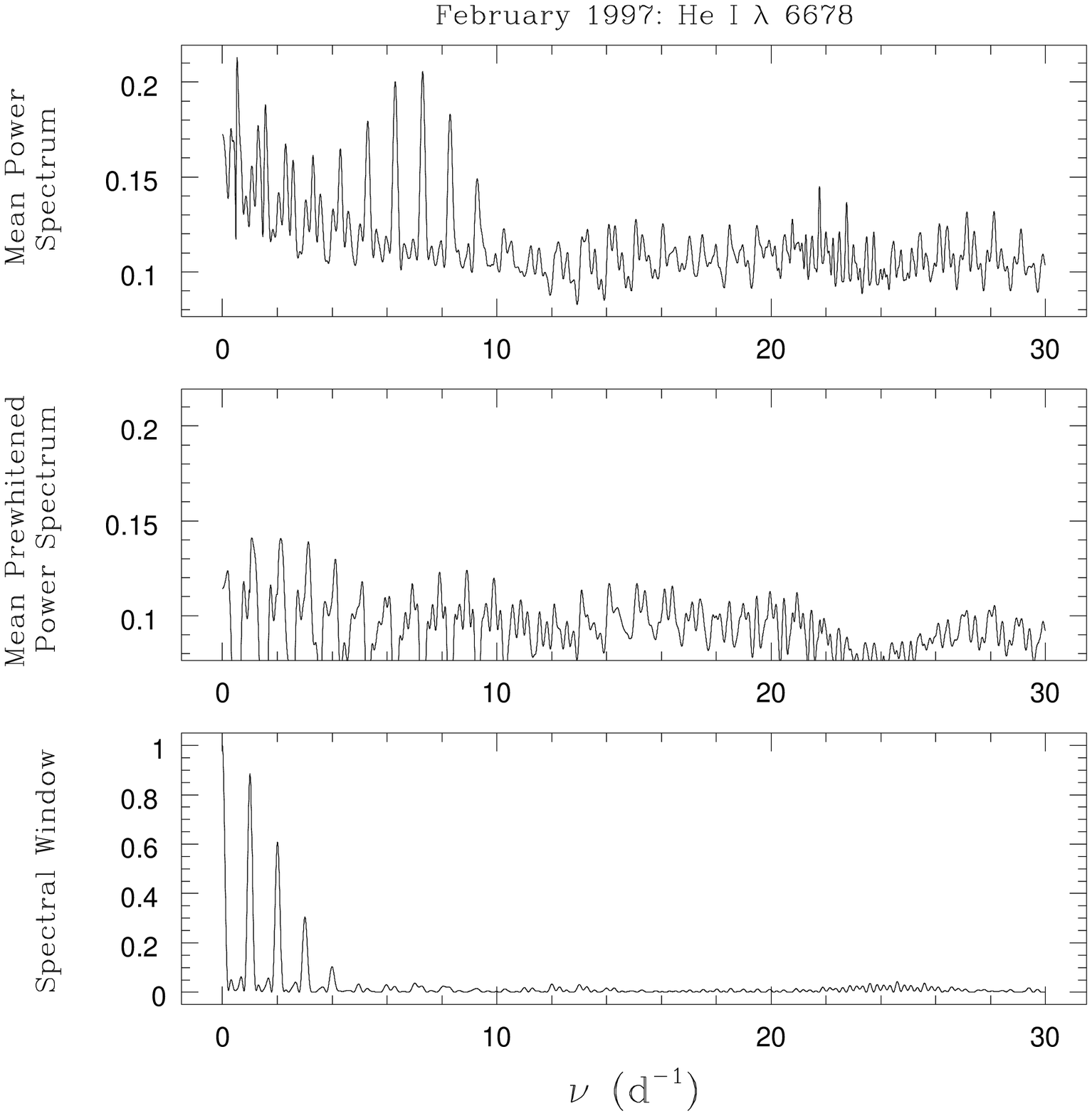}}
\end{minipage}
\hfill
\begin{minipage}{9.0cm}
\resizebox{9.0cm}{!}{\includegraphics{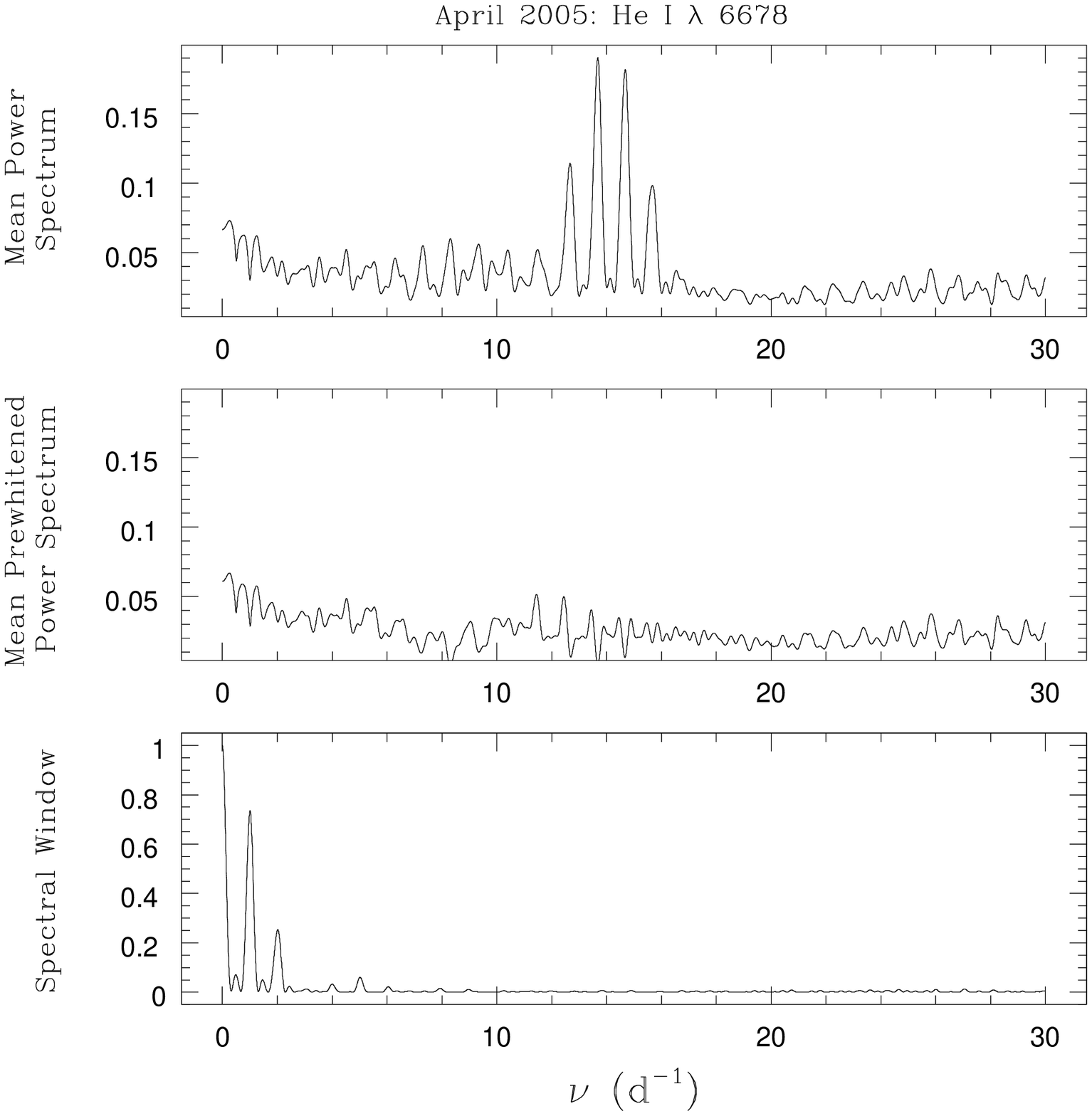}}
\end{minipage}
\caption{Left: the top panel yields the mean Fourier power spectrum of the February 1997 He\,{\sc i} $\lambda$\,6678 data as derived from the 2-D Fourier spectrum averaged over the profile of the He\,{\sc i} $\lambda$\,6678 line (from 6660 to 6700\,\AA). The middle panel indicates the mean power spectrum prewhitened for the two dominant frequencies (0.54 and 7.30\,d$^{-1}$) simultaneously. The lower panel yields the power spectral window resulting from the sampling of the February 1997 time series. Right: same for the April 2005 data. The frequencies used for the prewhitening are $\nu_1 = 13.68$ and $\nu_2 =  8.31$\,d$^{-1}$ in this case.\label{fourmean6678}}
\end{figure*}

We have determined the star's projected rotational velocity $v\,\sin{i}$ applying the Fourier method (see Gray \cite{Gray}, Sim\'on-D\'{\i}az \& Herrero \cite{SDH}) to the mean profiles of the He\,{\sc i} $\lambda\lambda$\,5876, 6678 and O\,{\sc iii} $\lambda$\,5592 lines. To first order, the structure of the Fourier transforms of these lines can be explained by projected equatorial rotational velocities of 345, 395 and 380\,km\,s$^{-1}$ respectively. While the results for the He\,{\sc i} $\lambda$\,6678 and O\,{\sc iii} $\lambda$\,5592 line are in very reasonable agreement, He\,{\sc i} $\lambda$\,5876 yields a significantly lower value. There are several reasons why the three lines do not yield exactly the same result. First of all, the profile of He\,{\sc i} $\lambda$\,5876 is affected by emission that could fill-in the absorption in the wings of the line, hence leading to an apparently lower $v\,\sin{i}$. In addition, although we have used the mean profiles evaluated from a large number of observations, the line profile variability (see below) leaves some residual structure in the profiles. These features lead in turn to a Fourier transform that deviates from the one expected for a pure rotational broadening\footnote{Macroturbulence also affects the shape of the Fourier transform, but this is mainly an issue for very slow rotators which is not the case of HD\,93521.}. The cleanest result is obtained for the He\,{\sc i} $\lambda$\,6678 line (see Fig.\,\ref{vsini}). In summary, we conclude that the projected equatorial rotational velocity of the star is most likely $390 \pm 10$\,km\,s$^{-1}$. There is a caveat here: our determination of $v\,\sin{i}$ likely represents a lower limit to the actual value. Indeed, Townsend et al.\ (\cite{TOH}) argued that the rotational velocities of very rapidly rotating Be-type stars might be systematically underestimated as a result of the effect of gravity darkening that would reduce the weight of the equatorial region (and hence of the highest rotational velocities) in the formation of the absorption line profiles. For B0 stars, Townsend et al.\ find that the He\,{\sc i} $\lambda$\,4471 and  Mg\,{\sc ii} $\lambda$\,4481 lines underestimate $v\,\sin{i}$ by about 10\%. A similar underestimate could hold in our case.\\ 

To quantify the variability of the spectral lines, we have computed the temporal variance spectrum (TVS, Fullerton et al.\ \cite{FGB}) for the data of each observing campaign. The TVS diagrams of the spectral regions investigated here reveal no significant variability in the metallic lines, except perhaps for some marginal variations in the region between 5660 and 5810\,\AA. The significant variability is restricted to the He\,{\sc i} $\lambda$\,4471, H$\beta$, He\,{\sc i} $\lambda$\,5876, H$\alpha$ and He\,{\sc i} $\lambda$\,6678 lines. The majority of these lines (except perhaps the H$\beta$ line, see Fig.\,\ref{average}) exhibit evidence for emission wings.

\subsection{Fourier analysis\label{sect: fourier}}
For each of the variable lines, we have performed a 2-D Fourier analysis (see e.g.\ Rauw et al.\ \cite{R01,R03}) where the Fourier power spectrum is calculated at each wavelength step across the line profile using the technique of Heck et al.\ (\cite{HMM}) revised by Gosset et al.\ (\cite{wr30a}). This algorithm is specifically designed to handle time series with an uneven sampling. The periodograms were calculated for each line and for each campaign up to a maximum frequency of 30\,d$^{-1}$. 

For the frequencies that are identified from the periodograms, we have further applied the different techniques outlined in Rauw et al.\ (\cite{R01}). For each line and for each observing campaign, we have fitted an expression of the type 
\begin{equation}
I(\lambda, t) = I_0(\lambda) + \sum_{i=1}^{q} A_i(\lambda)\,\sin{[2\,\pi\,\nu_i\,t + \phi_i(\lambda)]}
\label{prewhite}
\end{equation}
to the time series. Here $I(\lambda, t)$ is the line intensity at wavelength $\lambda$ and at time $t$, $\nu_i$ are the frequencies corresponding to the highest peaks in the periodogram, whilst $A_i(\lambda)$ and $\phi_i(\lambda)$ are respectively the semiamplitudes and the phases for these frequencies as a function of wavelength (see Rauw et al.\ \cite{R01})\footnote{These techniques are especially useful if the line profile variations are interpreted in terms of non-radial pulsations. Indeed, in this case, Telting \& Schrijvers (\cite{TS}) and Schrijvers \& Telting (\cite{ST}) have shown that the observable phase differences between the blue and red line wings can be related directly to the degree $l$ and to the absolute value of the azimuthal order $|m|$ of the pulsation mode.}. The number $q$ of frequencies that were simultaneously fitted was progressively increased from 1 up to 3. This method not only allows us to characterize the properties of a given frequency, but further enables us to `prewhiten' the data by removing the modulation at this frequency from the time series and re-compute the periodogram for the `prewhitened' data set. We start by removing the highest peak frequency found in the periodogram of the original data set. Next, we identify the highest frequency in the periodogram of the new (prewhitened) time series and prewhiten then the data simultaneously for the two frequencies. The procedure is repeated for up to three different frequencies\footnote{We caution that some of the low frequencies identified throughout this work likely correspond to long-term trends rather than to actual periods.}. Indeed, three frequencies are usually sufficient to reduce the residual amplitude in the prewhitened Fourier power spectra to a level compatible with the noise level of the data.

To evaluate the error bars on the semiamplitudes and phases of the various Fourier components, we have used Monte Carlo simulations assuming that the uncertainties on $I(\lambda, t)$ are equal to the noise level $\sigma(t)$ of the corresponding spectrum  evaluated in nearby continuum windows. In a subsequent step, we have also included the first harmonics of the three most significant frequencies (hence accounting for six frequencies in Eq.\,\ref{prewhite}). Usually however, no significant power was found in the first harmonics of those frequencies that are likely associated with pulsations (see below). The amplitudes of these harmonics are found to be very low, much lower than those of the actual frequencies. The resulting uncertainties on the phase constants of the harmonics are very large, rendering any attempt to measure their blue to red phase shift hopeless.  

The evaluation of the absolute significance of a peak in a periodogram of a time series with an uneven sampling is a controversial problem, that does not have an exact solution (see Heck et al.\ \cite{HMM}). Therefore, as a first order indication of the significance of a peak, we have evaluated, at each wavelength step $\lambda$, the quantity $z_{\nu}(\lambda) = P_{\nu}(\lambda)/s_{\nu}(\lambda)^2$ where $P_{\nu}(\lambda)$ is the power in the periodogram at the frequency $\nu$ under investigation and $s_{\nu}(\lambda)^2$ is the variance of the time series after prewhitening of this frequency. Whilst $z_{\nu}(\lambda)$ is not constant over the width of the various lines studied here, all the frequencies considered in the following yield rather large ($> 20$) values of $z_{\nu}(\lambda)$ over a significant part of the line profile. Therefore, all the frequencies discussed below are detected with a good significance. However, this does not imply that all the frequencies reported below actually have a physical meaning. Some of them, especially the low frequencies, could rather result from transient features in the wind of the star. 

In the following, we shall refer to $\nu_1$ and $\nu_2$ as the 13.6 and 8.3\,d$^{-1}$ frequencies previously reported in the literature.

\subsubsection{He\,{\sc i} $\lambda$\,6678}
The power spectrum calculated from the February 1997 time series of this line reveals its highest peak at 0.54\,d$^{-1}$ (see Fig.\,\ref{fourmean6678}). Beside this low frequency peak, the periodogram exhibits a family of aliases associated with $7.30$\,d$^{-1}$. In the raw periodogram, the one-day alias at 6.30\,d$^{-1}$ is stronger than the one at 8.29\,d$^{-1}$. However, this situation changes when the data are prewhitened for the 0.54\,d$^{-1}$ modulation with the 8.29\,d$^{-1}$ peak becoming stronger than the one at 6.30\,d$^{-1}$. 
\begin{figure}[h!]
\resizebox{9.0cm}{!}{\includegraphics{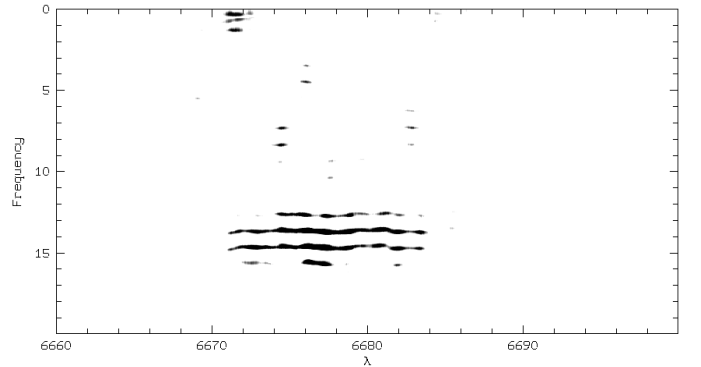}}
\caption{Grey-scale 2-D Fourier power spectrum of the He\,{\sc i} $\lambda$\,6678 line as derived from the April 2005 data. \label{6678april2005}}
\end{figure}

The lower frequency apparently produces a modulation of the entire line profile; there is no significant change of the phase as a function of wavelength. This variation is therefore unlikely to be related to non-radial pulsations. On the other hand, the 7.30\,d$^{-1}$ frequency is an alias of the $\nu_2 = 8.28$\,d$^{-1}$ frequency reported by Howarth et al.\ (\cite{HTC}). We have prewhitened the data using either the 7.30\,d$^{-1}$ or the 8.29\,d$^{-1}$ frequencies. Due to the strong level of the noise `continuum' in the February 1997 periodogram, the prewhitening procedure does not result in a clean periodogram. Still, the results obtained when prewhitening either of the alias frequencies are the same within the level of the noise in the periodogram. Both candidate frequencies display a progression of the phase constant $\phi_i$ from the blue wing to the red wing. This progression is not fully monotonic (at least partially because of the noise level of the data). Nevertheless, assuming $v_e\,\sin{i} = 395$\,km\,s$^{-1}$, we find that the blue -- red phase difference $\Delta\,\phi$ amounts to about $3\,\pi$ radians, irrespective of the choice of the alias. However, we stress that the amplitude associated with $\nu_2$ is largest over the radial velocity interval $-350$ to $+100$\,km\,s$^{-1}$: there is almost no significant signal outside this velocity domain (i.e.\ between $+100$\,km\,s$^{-1}$ and $+v_e\,\sin{i}$). 
\begin{figure}[htb!]
\resizebox{9.0cm}{!}{\includegraphics{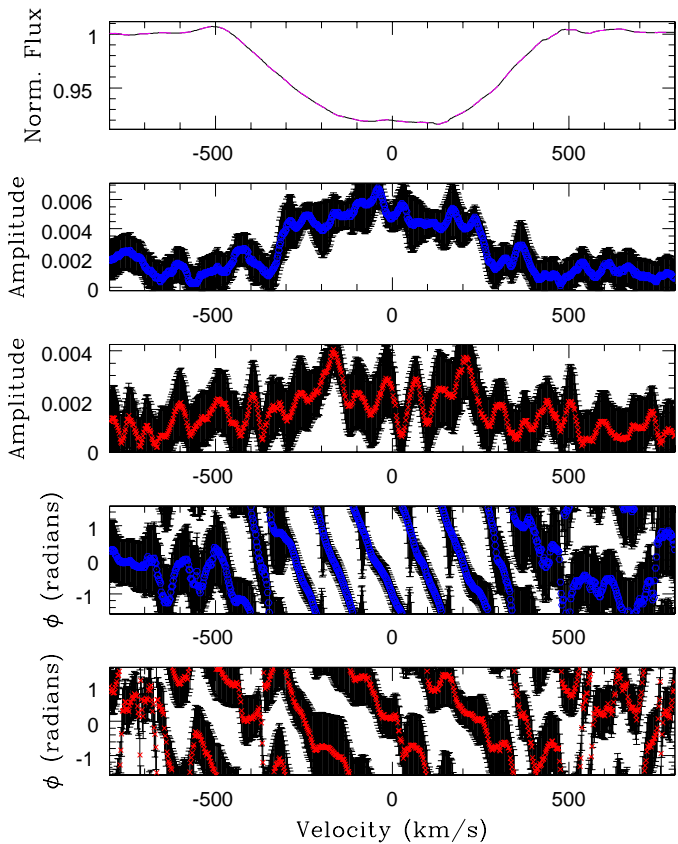}}
\caption{The top panel yields the mean normalized He\,{\sc i} $\lambda$\,6678 line profile as observed in April 2005 shown as a continuous line, as well as the mean profile after prewhitening (i.e.\ the $I_0(\lambda)$ term in Eq.\,\ref{prewhite}) shown by a dashed line. The next panels (from top to bottom) yield the semiamplitudes (in units of the continuum level) of the $\nu_1$ (dots, second panel) and $\nu_2$ (crosses, third panel) modulations, the phase constant $\phi_1$ of the $\nu_1$ signal and the phase constant $\phi_2$ of the $\nu_2$ component. The error bars were obtained from Monte Carlo simulations (see text). Phase 0.0 was arbitrarily set to the time of the first observation of this data set.\label{phase6678april2005}}
\end{figure}

\begin{figure*}[htb]
\begin{minipage}{9.0cm}
\resizebox{9.0cm}{!}{\includegraphics{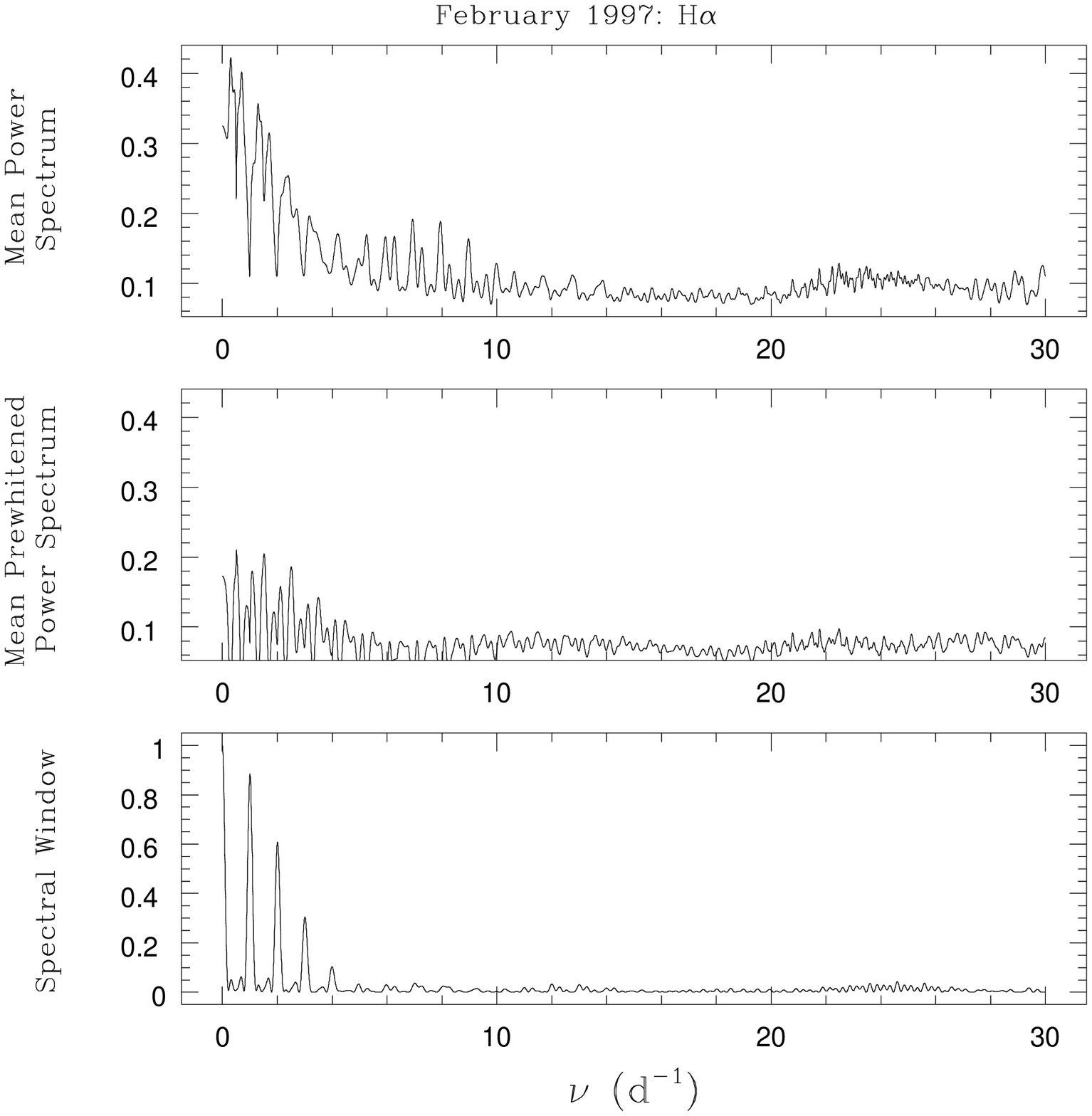}}
\end{minipage}
\hfill
\begin{minipage}{9.0cm}
\resizebox{9.0cm}{!}{\includegraphics{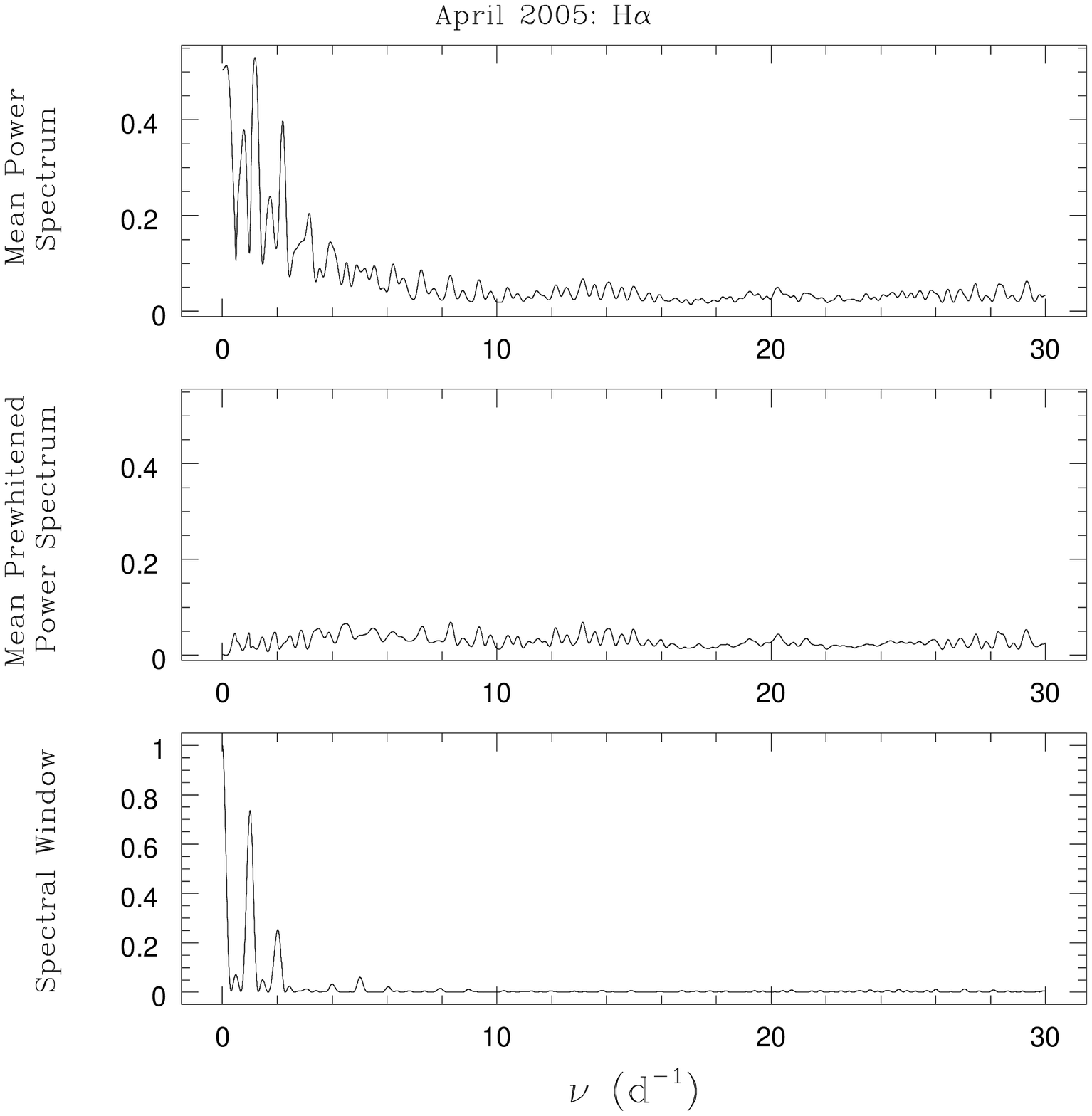}}
\end{minipage}
\caption{Same as Fig.\,\ref{fourmean6678} but for the H$\alpha$ line: February 1997 (left) and OHP April 2005 (right). The power spectra are averaged over the wavelength domain from 6540 to 6580\,\AA. The frequencies used for prewhitening are 0.30, 6.94 and 6.26\,d$^{-1}$ for the February 1997 data and 1.18, 3.09 and 1.67\,d$^{-1}$ for the April 2005 campaign. \label{fourmeanHa}}
\end{figure*}

The power spectrum of the April 2005 data looks completely different (Fig.\,\ref{fourmean6678}). The periodogram is now dominated by a strong peak at $\nu_1 = 13.68$\,d$^{-1}$ along with its one-day aliases. When the data are prewhitened for the $\nu_1$ frequency, another family of aliases associated with $\nu_2 = 8.31$\,d$^{-1}$ becomes dominant. Again, the mean power spectrum, averaged between 6660 and 6700\,\AA, reveals some power also at lower frequencies (near 0.25\,d$^{-1}$), though the 2-D power spectrum indicates that this latter feature is mainly restricted to a narrow range of wavelengths in the line wings and mostly in the blue wing near 6671\,\AA\ (see Fig.\,\ref{6678april2005}). 

Although the power spectrum reveals some structure near 26\,d$^{-1}$, the frequency of the strongest peak in this region (25.8\,d$^{-1}$) is quite different from $2 \times \nu_1$ (= 27.36\,d$^{-1}$) and it seems therefore unlikely that this feature is related to the first harmonic of $\nu_1$. The periods corresponding to $\nu_1$ and $\nu_2$ are respectively 1.75 and 2.89\,hr, in excellent agreement with the periods ($P_1 = 1.76 \pm 0.04$\,hr, $P_2 = 2.90 \pm 0.12$\,hr) reported by Howarth et al.\ (\cite{HTC}). 

Figure\,\ref{phase6678april2005} illustrates the mean line profile as observed in April 2005, as well as the semiamplitudes and phases of the two dominant frequencies ($\nu_1$ and $\nu_2$) during this campaign. Again, we note the diagonal progression of the phase constant as a function of wavelength which is typical of non-radial pulsations. From Fig.\,\ref{phase6678april2005}, we infer a blue to red phase difference of $\Delta\,\phi_1 = 7\,\pi$ and $\Delta\,\phi_2 \simeq 3.5\,\pi$ for the $\nu_1$ and $\nu_2$ frequencies respectively, although for $\nu_2$, the progression of the phase constant is not strictly monotonic across the line profile. 

\subsubsection{H$\alpha$ and H$\beta$}
The power spectrum of the February 1997 H$\alpha$ data shows the highest peak at $0.30$\,d$^{-1}$ and its one-day alias at 0.70\,d$^{-1}$. A group of secondary peaks is found at 6.94\,d$^{-1}$ and its aliases. We further detect some weaker peaks at 6.26\,d$^{-1}$ (alias of $\nu_2$) as well as its aliases. The power spectrum is rather well cleaned by prewhitening these three frequencies (see Fig.\,\ref{fourmeanHa}). The low frequency variation corresponds to a global modulation of the line intensity in the core (radial velocity interval from $-100$ to $+100$\,km\,s$^{-1}$) with a rather constant phase. Since the 6.26\,d$^{-1}$ frequency is an alias of $\nu_2$, we have also prewhitened the data using 0.30, 6.94 and 8.28\,d$^{-1}$. For the latter frequency, we observe again a blue to red phase difference of about $3\,\pi$ (within the uncertainties). However, as for He\,{\sc i} $\lambda$\,6678, the amplitude of this modulation is really significant only in the blue wing of the line. 

In the April 2005 OHP H$\alpha$ data, the highest peak of the power spectrum occurs at $0.15$\,d$^{-1}$ and its aliases (1.18 and 0.78\,d$^{-1}$). Prewhitening either of these frequencies yields a set of secondary peaks near 1.67 and 3.09\,d$^{-1}$. If these frequencies are removed in turn, the power spectrum of the residual data set displays some low amplitude peaks at 8.32 and 13.14\,d$^{-1}$ (note that while the former frequency is likely equal to $\nu_2$, the latter one is formally not consistent with $\nu_1$). 

The April 2005 SPM echelle spectra yield a similar picture: the bulk of the power of the H$\alpha$ variations is again concentrated in variations that occur on low-frequency timescales (1.09 and 1.32\,d$^{-1}$, which are also found for the He\,{\sc i} $\lambda$\,5876 line in the same data set) and affect mainly the wings of the line. Prewhitening the data for these two frequencies yields a highest peak at 2.51\,d$^{-1}$ as well as a series of aliases of $\nu_1 + 1$ and $\nu_2 - 1$. For H$\beta$, the SPM data reveal the strongest peaks at 0.01 and 0.44\,d$^{-1}$. Both, $\nu_2$ and $\nu_1$ are clearly detected though and the blue to red phase differences are about $3\,\pi$ and $7\,\pi$, though it should be stressed here that the evolution of the phase constants is not monotonic in the case of this line.

\begin{figure*}[thb!]
\begin{minipage}{8.5cm}
\resizebox{8.5cm}{!}{\includegraphics{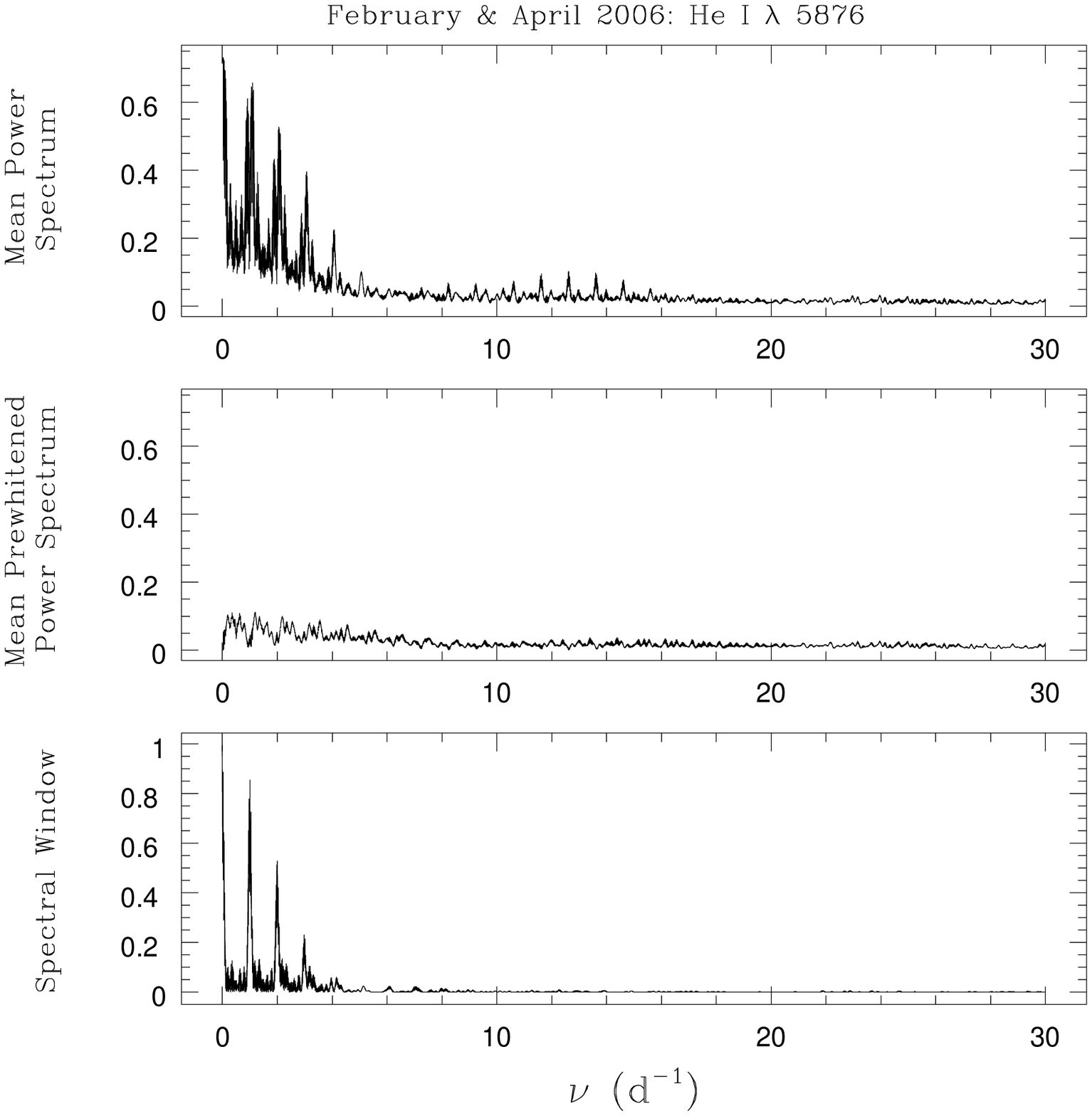}}
\end{minipage}
\hfill
\begin{minipage}{8.0cm}
\resizebox{8.0cm}{!}{\includegraphics{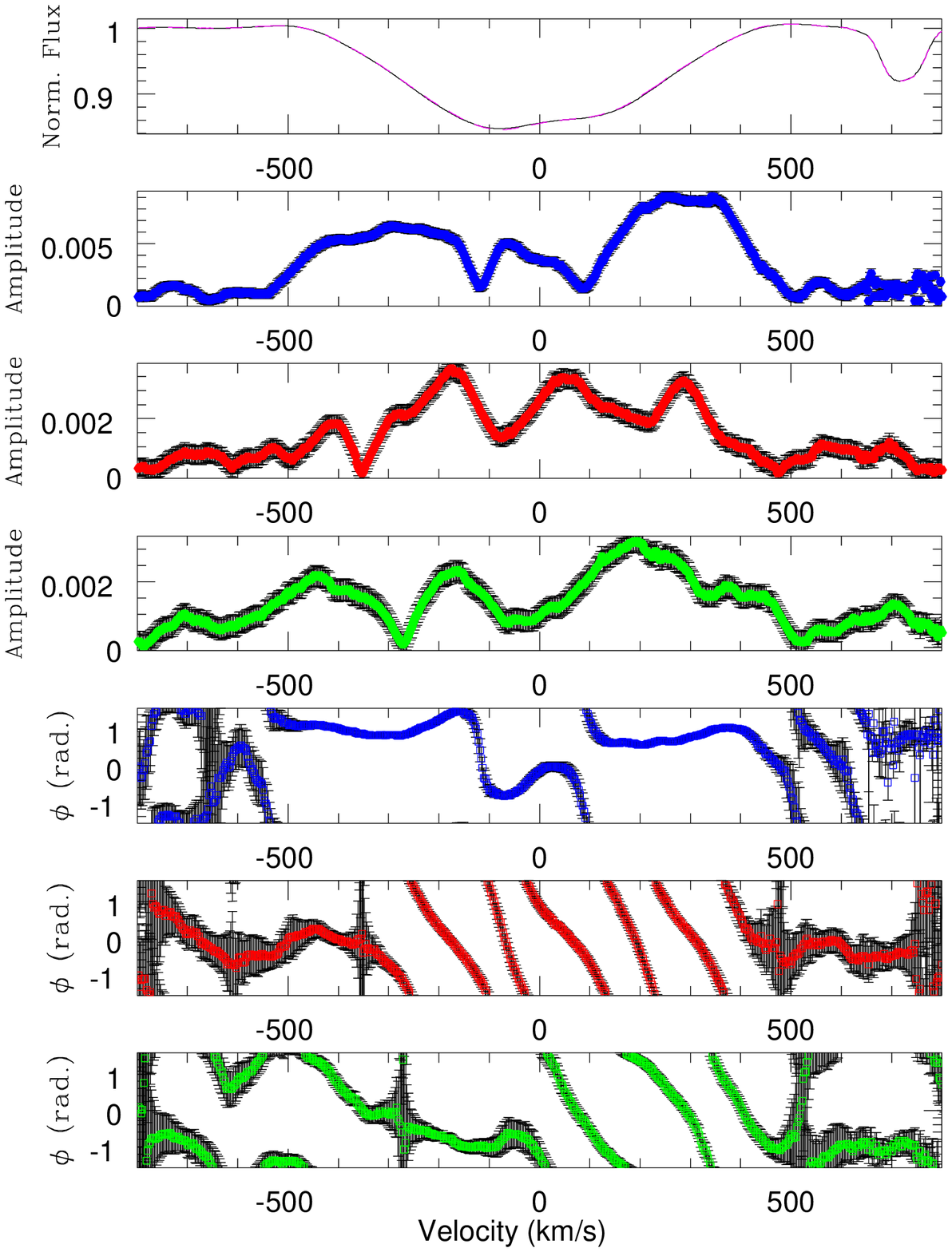}}
\end{minipage}
\caption{Left: same as Fig.\,\ref{fourmean6678} but for the He\,{\sc i} $\lambda$\,5876 line as observed in February and April 2006. The power spectra are averaged over the wavelength domain from 5860 to 5900\,\AA. The frequencies used for prewhitening are 1.10, 12.63 and 8.26\,d$^{-1}$. Right: same as Fig.\,\ref{phase6678april2005} but for the He\,{\sc i} $\lambda$\,5876 line as observed in February and April 2006. The semiamplitudes and phases of the 1.10, 13.60 and 8.26\,d$^{-1}$ frequencies are shown from top to bottom in the right panel.\label{fourmean5876}}
\vspace*{5mm}
\begin{minipage}{8.5cm}
\resizebox{8.5cm}{!}{\includegraphics{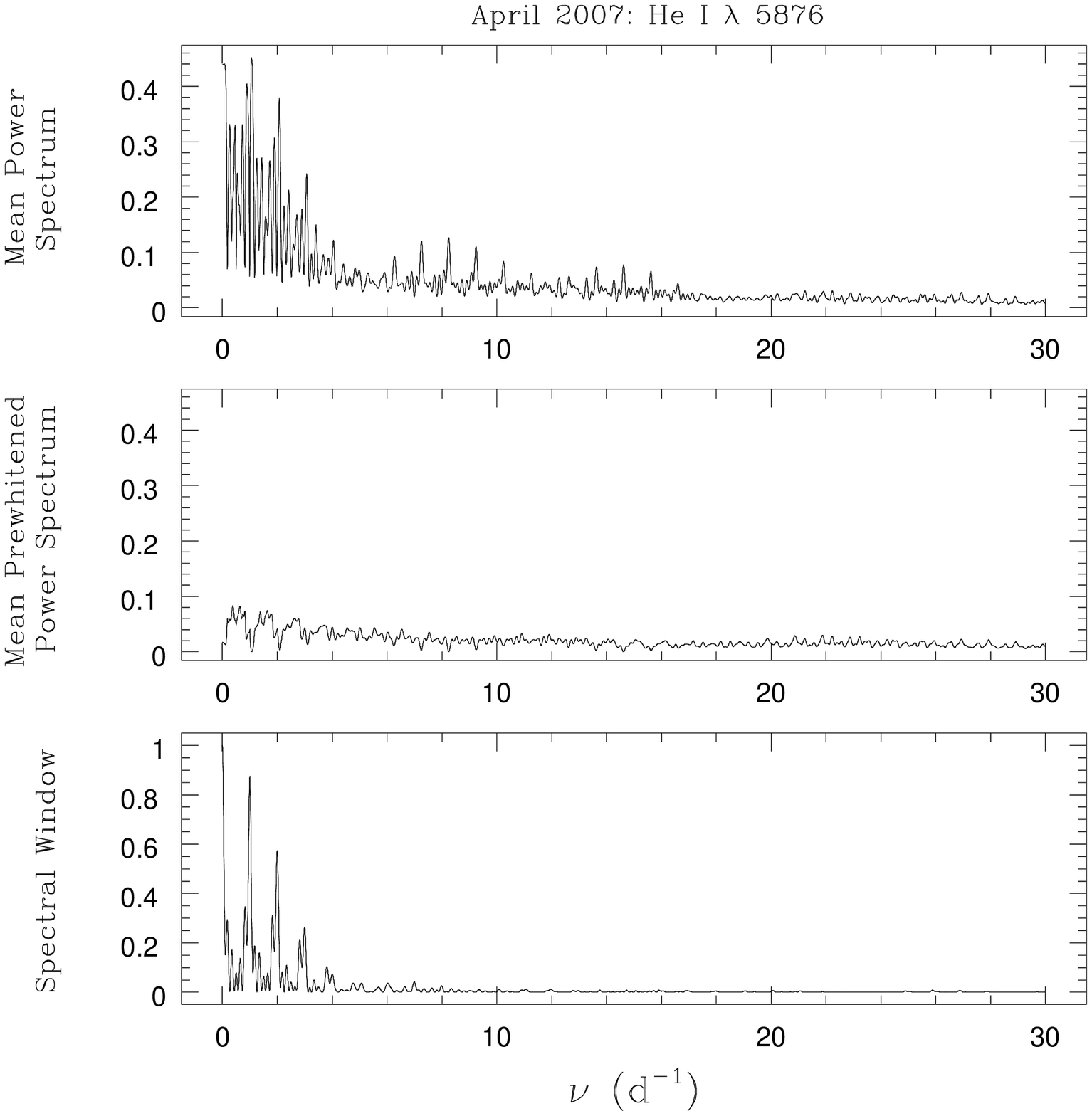}}
\end{minipage}
\hfill
\begin{minipage}{8.0cm}
\resizebox{8.0cm}{!}{\includegraphics{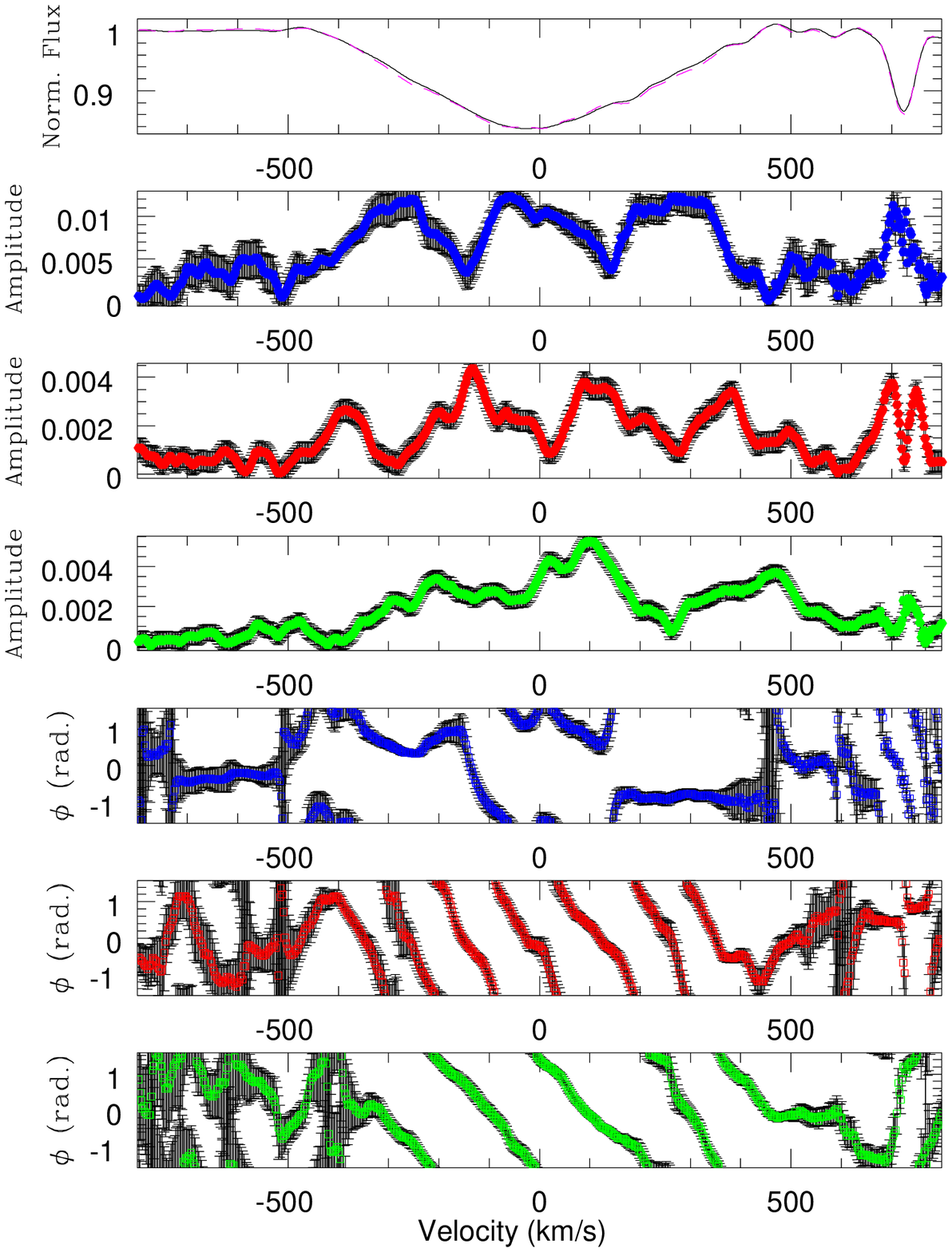}}
\end{minipage}
\caption{Same as Fig.\,\ref{fourmean5876} but for the He\,{\sc i} $\lambda$\,5876 line as observed in April 2007. The frequencies used for prewhitening are 1.06, 14.62 and 8.25\,d$^{-1}$. \label{four5876}}
\end{figure*}

\subsubsection{He\,{\sc i} $\lambda\lambda$\,5876, 4471}
The power spectrum of the OHP 2006 data is dominated by low frequencies. When all the OHP data taken in February and April 2006 are analysed together, the highest peak is found at $0.02$\,d$^{-1}$ (along with its aliases, Fig.\,\ref{fourmean5876}). Prewhitening the time series for this frequency and its first harmonics, yields however some artefacts in the mean line profile ($I_0(\lambda)$ in Eq.\,\ref{prewhite}). Hence, we rather used the 1-day alias $1.10$\,d$^{-1}$. When we prewhiten the data for the latter frequency, the highest peaks in the new periodogram are found at 12.63 and 8.26\,d$^{-1}$ ($= \nu_2$). The former frequency is likely an alias of $\nu_1$. Prewhitening the frequencies 1.10, 8.26 and 13.60\,d$^{-1}$, we find that the blue to red phase difference amounts to $6\,\pi$ for $\nu_1$ (for 12.63\,d$^{-1}$, this becomes $\simeq 7\,\pi$). For $\nu_2$ we find a phase difference of about $3.5\,\pi$ with the caveat that the phase progression is apparently not monotonic for this frequency (see Fig.\,\ref{fourmean5876}).
These results are largely confirmed by the analysis of the He\,{\sc i} $\lambda$\,5876 line in the April 2005 SPM time series. Whilst the variations in the wings of the line are dominated by two low frequencies at $1.08$ and $1.30$\,d$^{-1}$, $\nu_1$ (or its alias at 12.59\,d$^{-1}$) as well as $\nu_2$ are detected with a roughly constant amplitude over the entire width of the line profile and the blue to red phase differences amount to $\sim 7\,\pi$ and $\sim 4.5\,\pi$ for $\nu_1$ and $\nu_2$ respectively. 

The SPM echelle spectra also cover the He\,{\sc i} $\lambda$\,4471 line. The frequency content of the power spectrum of the latter essentially confirms the results obtained for the He\,{\sc i} $\lambda$\,5876 line: a dominant peak at $0.08$\,d$^{-1}$ (alias of the $1.08$\,d$^{-1}$ peak reported above), as well as two pulsation signals found at $\nu_1$ as well as $\nu_2-1$. The main difference with the above results concerns the blue to red phase differences which, in the case of the He\,{\sc i} $\lambda$\,4471 line, amount to $\sim 5.8\,\pi$ and $\sim 3\,\pi$ for $\nu_1$ and $\nu_2 - 1$ respectively. 

A similar description applies also to the April 2007 periodogram of the He\,{\sc i} $\lambda$\,5876 time series. Indeed, the highest peak is again found at low frequencies, $1.06$\,d$^{-1}$ (along with its aliases, Fig.\,\ref{four5876}). Prewhitening the time series for this frequency yields secondary peaks at 8.25 ($= \nu_2$) and 14.61\,d$^{-1}$ (likely the $\nu_1 + 1$ alias). Prewhitening the frequencies 1.06, 8.25 and 14.61\,d$^{-1}$, we find that the blue to red phase difference amounts to $6.5\,\pi$ for the 14.61\,d$^{-1}$ frequency (the same result holds if we adopt $\nu_1$ instead). For $\nu_2$ we find a phase difference of about $4\,\pi$ (see Fig.\,\ref{four5876}). 

\begin{table*}[thb!]
\caption{Results of the Fourier analysis of the photometric time series of HD\,93521. Columns 2, 3, 7, 8, 9 and 10 yield the average magnitude and the dispersion about the mean for HD\,93521 and the two comparison stars. Columns 4, 5 and 6 list the highest peaks in the periodogram of the photometric time series of HD\,93521. The numbers in brackets provide the semiamplitude of the variation associated with each of these frequencies. \label{fourphoto}}
\begin{tabular}{c c c c c c c c c c c c}
\hline
Filter & \multicolumn{5}{c}{HD\,93521} & & \multicolumn{2}{c}{HD\,90250} & & \multicolumn{2}{c}{HD\,96951} \\
\cline{2-6}\cline{8-9}\cline{11-12}
       & mag & $\sigma$ & \multicolumn{3}{c}{Frequencies} & & mag & $\sigma$ && mag & $\sigma$ \\
\hline
$V$   & 7.0043 & 0.0047 & 7.90\,d$^{-1}$ (2.9\,mmag)  & 2.05\,d$^{-1}$ (2.5\,mmag) &7.10\,d$^{-1}$ (2.3\,mmag) && 6.4900 & 0.0032 && 7.8702 & 0.0035 \\
$B$   & 5.7589 & 0.0053 & 0.88\,d$^{-1}$ (3.4\,mmag)  & 7.90\,d$^{-1}$ (3.2\,mmag) &6.11\,d$^{-1}$ (2.4\,mmag) && 6.8883 & 0.0045 && 6.9591 & 0.0038 \\
$U$   & 5.8809 & 0.0079 & 2.00\,d$^{-1}$ (11.2\,mmag) & 6.10\,d$^{-1}$ (3.5\,mmag) &7.90\,d$^{-1}$ (3.4\,mmag) && 9.0681 & 0.0085 && 8.4939 & 0.0060 \\
$B_1$ & 6.5113 & 0.0057 & 0.88\,d$^{-1}$ (4.9\,mmag)  & 4.03\,d$^{-1}$ (2.8\,mmag) &7.90\,d$^{-1}$ (2.6\,mmag) && 8.2339 & 0.0052 && 7.8505 & 0.0048 \\
$B_2$ & 7.3870 & 0.0057 & 0.88\,d$^{-1}$ (4.4\,mmag)  & 7.90\,d$^{-1}$ (4.3\,mmag) &6.81\,d$^{-1}$ (3.1\,mmag) && 8.0003 & 0.0039 && 8.4397 & 0.0047 \\
$V_1$ & 7.6856 & 0.0057 & 2.00\,d$^{-1}$ (5.4\,mmag)  & 0.88\,d$^{-1}$ (4.3\,mmag) &7.90\,d$^{-1}$ (3.0\,mmag) && 7.2655 & 0.0040 && 8.5661 & 0.0050 \\
$G$   & 8.2383 & 0.0056 & 0.04\,d$^{-1}$ (4.3\,mmag)  & 7.58\,d$^{-1}$ (2.8\,mmag) &4.93\,d$^{-1}$ (2.5\,mmag) && 7.4440 & 0.0036 && 9.0553 & 0.0055 \\
\hline
\end{tabular}
\end{table*}
\subsection{Photometry}
\begin{figure}[htb!]
\resizebox{9.0cm}{!}{\includegraphics{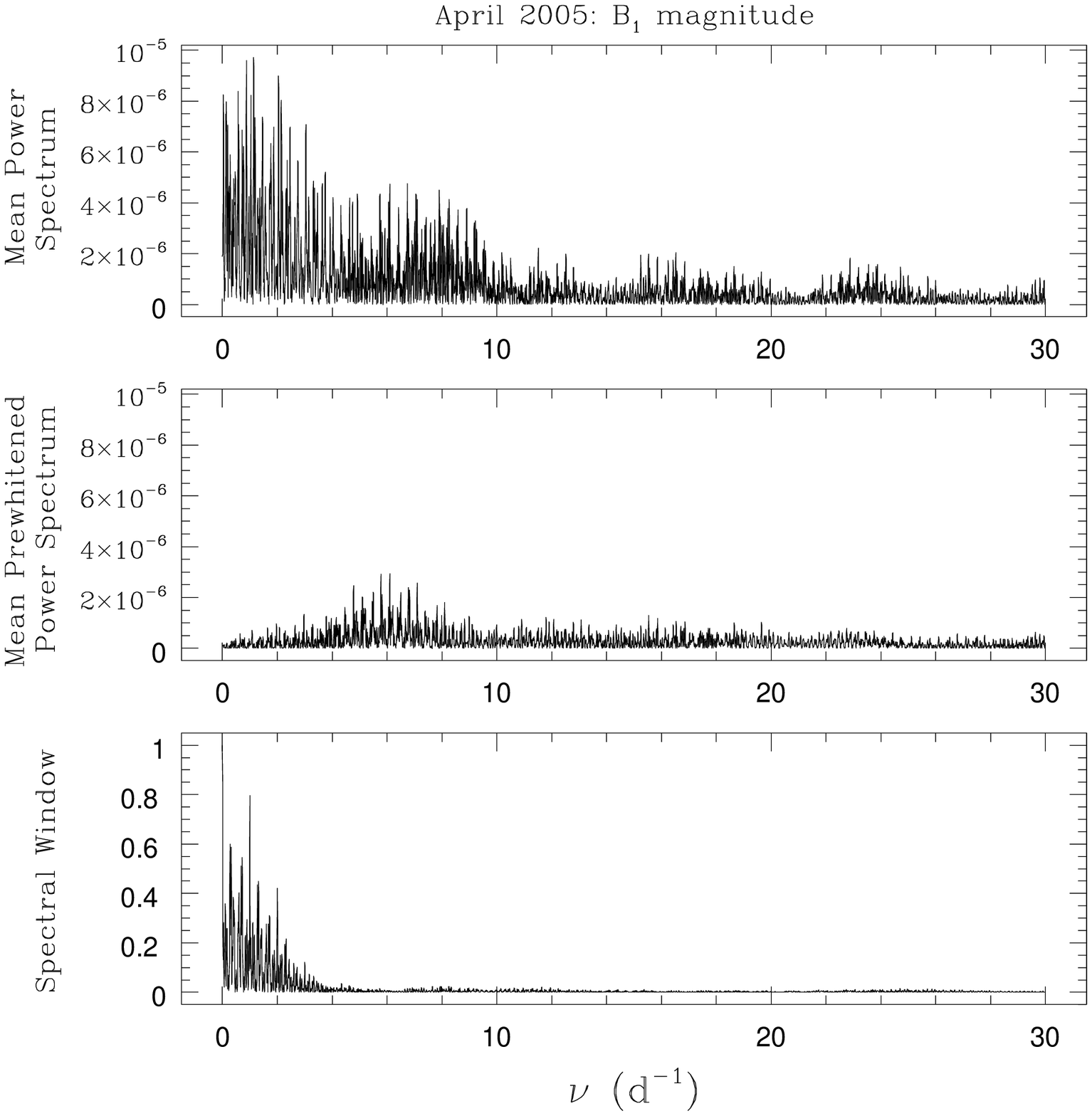}}
\caption{Fourier power spectrum of the $B_1$ data of HD\,93521. The top panel yields the raw power spectrum, whilst the middle panel corresponds to the power spectrum of the time series prewhitened for three frequencies (2.04, 6.74 and 8.54\,d$^{-1}$). The lower panel displays the spectral window of the photometric time series.\label{photom}} 
\end{figure}

We have analysed the time series of HD\,93521 in each of the seven filters of the Geneva system using the Fourier method of Heck et al.\ (\cite{HMM}). For each filter, we find that the periodogram indicates significant power at frequencies below about 10\,d$^{-1}$. In all cases, we find that the power spectrum can efficiently be prewhitened with three frequencies (see e.g.\ Fig.\,\ref{photom}).

Some frequencies are found consistently in several filters though with different amplitudes. This is the case for 7.90\,d$^{-1}$ (found in six filters out of seven), 0.88\,d$^{-1}$ (detected in four filters) and 2.00\,d$^{-1}$ (detected in three filters). None of the frequencies reported here is observed in the April 2005 spectroscopic time series that was obtained in coordination with the photometric campaign. 

The $U$ photometry yields somewhat different results from most other filters: the variations are clearly dominated by the 2.00\,d$^{-1}$ frequency, while in most other filters the various frequencies appear with similar amplitudes. At first sight, this might seem surprising because, for OB stars, the signatures of non-radial pulsations are usually expected to be quite strong in the $U$ filter (De Cat et al.\ \cite{DeCat}). We caution that frequencies close to an integer number of d$^{-1}$ (such as the 2.00\,d$^{-1}$ frequency) could actually be artefacts due to the interplay between the sampling of our time series and a few deviating data points. In this respect, we emphasize that the $U$ filter is most heavily affected by atmospheric absorption and therefore, the 2.00\,d$^{-1}$ signal could be related to imperfect extinction corrections, at least in the $U$ filter. 

Table\,\ref{fourphoto} also lists the mean magnitudes and the 1-$\sigma$ standard deviations of the measurements of HD\,93521 as well as of two reference stars (HD\,90250, K1\,III and HD\,96951, A1\,V) that were observed during the same campaign. The dispersion of the measurements of HD\,93521 is always larger than those of the reference stars including in the filters where HD\,93521 is significantly brighter than the reference stars. The only exception is the $U$ filter where the largest value of $\sigma$ is found for HD\,90250, however this star is by far the faintest in the $U$ band. We further note that the Fourier power spectra of both reference stars are quite different from those of HD\,93521 and are essentially consistent with white noise.   

In summary, HD\,93521 appears to be slightly variable in photometry, though the frequency content of the variations is rather complex and the frequencies found in the photometric time series have very small amplitudes. It is likely that an even longer time series of very accurate photometric measurements (maybe from a space-borne observatory) is needed to disentangle the frequency spectrum.
 
\begin{table*}[htb!]
\caption{Maximum semi-amplitudes (in units of the continuum) of the frequencies detected in the 2-D Fourier analyses of the line profile variability of HD\,93521. For each line and each epoch, the amplitudes are given for those frequencies that were detected with a good significance above the noise level.\label{tabampl}}
\begin{center}
\begin{tabular}{r c c c c c c c c c c c c c c}
\hline
$\nu$ (d$^{-1}$) &  & He\,{\sc i} $\lambda$\,4471 & H$\beta$ & \multicolumn{3}{c}{He\,{\sc i} $\lambda$\,5876} & & \multicolumn{2}{c}{H$\alpha$} & & \multicolumn{2}{c}{He\,{\sc i} $\lambda$\,6678}\\
\cline{5-7} \cline{9-10} \cline{12-13}
& & SPM04/05 & SPM04/05 & SPM04/05 & 2006 & 04/07 & & 02/97 & 04/05 & & 02/97 & 04/05 \\
\hline
14.61 & $\sim 1 + \nu_1$ & & & & & 0.004 & & & & & & \\
13.68 & $\nu_1$ & 0.007 & 0.004 & 0.006 & & & & & & & & 0.006 \\
12.63 & $\sim \nu_1 - 1$ & & & & 0.004\\
8.31 & $\nu_2$ & & 0.005 & 0.004 & 0.003 & 0.005 & & & & & & 0.004\\
7.30 & $\sim  \nu_2 - 1$ & 0.007 & & & & & & & & & $\sim 0.008$ & \\
6.94 & & & & & & & & $\sim 0.008$ & & & & \\
6.26 & $\sim \nu_2 - 2$ & & & & & & & $\sim 0.006$ & & & & \\
3.09 & & & & & & & & & 0.008 & \\
1.67 & & & & & & & & & 0.008 & \\
1.30 & & & & 0.008 \\
1.08 & & & & 0.014 & & 0.011 \\
0.54 & & & & & & & & & & & $\sim 0.020$ & \\
0.44 & & & 0.006 \\
0.30 & & & & & & & & 0.015 & & & & \\
0.25 & & & & & & & & & & & & 0.004 \\
0.15 & & & & & & & & & 0.011 & \\
0.08 & & 0.012 \\
0.02 & & & & & 0.023\\
\hline
\end{tabular}
\end{center}
\end{table*}

\section{Discussion}
\subsection{General properties}
A summary of the amplitudes of all the frequencies detected in the spectroscopic part of this study is provided in Table\,\ref{tabampl}.

A first conclusion from this table is that the $\nu_1$ frequency is not always detected: while it clearly dominates in the April 2005 data of the He\,{\sc i} $\lambda$\,6678 line, it is absent from our February 1997 time series of the same line. Also, this frequency is never detected in the H$\alpha$ line profile variations. In the He\,{\sc i} $\lambda\lambda$\,5876 and 6678 lines, this frequency is associated with a maximum semiamplitude of modulation of 0.004 -- 0.006 in units of the continuum. On the other hand, $\nu_2$ or its aliases are detected in each line and each observing campaign. The maximum semiamplitudes of this mode are 0.003 (He\,{\sc i} $\lambda\lambda$\,5876, 6678 in 2006 and April 2005), 0.006 (H$\alpha$, February 1997) and 0.008 (He\,{\sc i} $\lambda$\,6678, February 1997) in units of the continuum. 

Another conclusion is the lack of any significant signal at the $\nu_3 = 2.66$\,d$^{-1}$ frequency ($P_3 = 9.0 \pm 1.2$\,hr) that was reported by Howarth et al.\ (\cite{HTC}) from their analysis of 103 {\it IUE} spectra with a median sampling of 0.60\,hr. This frequency is thus apparently absent from our data, except perhaps for the detection of the 2.51\,d$^{-1}$ signal in the April 2005 SPM H$\alpha$ data. 

Generally speaking, the (low-level) periodic variations with frequencies $\nu_1$ and $\nu_2$ are not the dominant source of line profile variations in any of the lines investigated here (except for the He\,{\sc i} $\lambda$\,6678 line in April 2005). In the majority of the cases, the most important variations are actually characterized by low frequencies. These low-frequency variations affect both the core of the lines as well as the wings and emission lobes (in the case of the H$\alpha$ and He\,{\sc i} $\lambda$\,5876 lines). Apparently, these modulations do not occur with a single stable clock: none of the low frequencies is detected in more than one line and for more than one observing campaign (except perhaps for the frequency around 1.10\,d$^{-1}$). It seems thus more appropriate to talk about time scales than periods for these longer term variations. This result casts doubt on a rotational modulation as the origin of the long-term variations: Howarth \& Reid (\cite{HR}) reported the analysis of 21 optical echelle spectra of HD\,93521 acquired over two nights in February 1992. From the variations of the emission wings of the H$\alpha$ and He\,{\sc i} $\lambda$\,5876 lines, these authors estimated a rotational period of 35\,hr ($\nu = 0.69$\,d$^{-1}$). None of the lines investigated here displays a significant signal at this frequency. The closest detections are found for the 1997 OHP H$\alpha$ data (the alias of the highest peak at 0.70\,d$^{-1}$) and for the same line observed in 2005 (the alias of the highest peak at 0.78\,d$^{-1}$). Therefore, our data do not confirm the existence of a rotational period of about 35\,hr and no unambiguous rotational period can be identified observationally for this star. 

In the remaining subsections we will focus on the interpretation of the variations seen at the $\nu_1$ and $\nu_2$ frequencies.
\subsection{$\nu_1$ and $\nu_2$ as non-radial pulsations}
In this section, we assume that the line profile variability at the frequencies $\nu_1$ and $\nu_2$ is due to two different non-radial pulsation modes. For multi-mode pulsations, one expects to observe variability with the genuine pulsation frequencies and their harmonics, as well as with their sums and beat frequencies (see e.g.\ the case of the O9.5\,V star $\zeta$\,Oph, Kambe et al.\ \cite{Kambe}, Walker et al.\ \cite{Walker}). In HD\,93521, we find no significant signal at the first harmonics of $\nu_1$ and $\nu_2$ nor at the sum or beat frequencies. Schrijvers \& Telting (\cite{ST}) argue that for non-radial pulsations with an amplitude of order 10\% of the mean line depth, the absence of a first harmonic is an indication that the line profile variability is mainly due to temperature effects (rather than to the Doppler-redistribution of flux). In HD\,93521, the amplitudes of the $\nu_1$ and $\nu_2$ modes in the He\,{\sc i} $\lambda$\,6678 line are of order 5 -- 10\% of the maximum line strength. While these amplitudes might be somewhat low for the harmonics to be detected, we nevertheless note that temperature effects could play a significant role in the line profile variability observed in HD\,93521.

\begin{table*}[htb!]
\caption{Properties of the frequencies detected in the 2-D Fourier analyses of the line profile variability of HD\,93521. The $l$ values were derived from the blue-to-red phase differences discussed in Sect.\,\ref{sect: fourier} using the formula of Telting \& Schrijvers (\cite{TS}).\label{tabspec}}
\begin{center}
\begin{tabular}{r c c c c l}
\hline
$\nu$ (d$^{-1}$) & P (hr) & Detection & $\Delta\,\phi/\pi$ & $l$ & Comment \\
\hline
14.61 & 1.64 & He\,{\sc i} $\lambda$\,5876 (2007) & 6.5 & 7.2 & likely alias of $\nu_1$ \\
13.68 & 1.75 & He\,{\sc i} $\lambda$\,6678 (04/05) & 7.0 & 7.7 & $= \nu_1$ \\
12.63 & 1.90 & He\,{\sc i} $\lambda$\,5876 (2006) & 7.0 & 7.7 & likely alias of $\nu_1$ \\
 8.31 & 2.89 & He\,{\sc i} $\lambda$\,6678, He\,{\sc i} $\lambda$\,5876 (2006,2007) & 4,3.5 & 4.5,3.9 & $= \nu_2$ \\
 7.30 & 3.29 & He\,{\sc i} $\lambda$\,6678 (02/97) & 3.0 & 3.4 & likely alias of $\nu_2$ \\
\hline
\end{tabular}
\end{center}
\end{table*}

Telting \& Schrijvers (\cite{TS}) and Schrijvers \& Telting (\cite{ST}) derived linear formulae relating the observable phase differences between the blue and red line wings to the degree $l$ and the absolute value of the azimuthal order $|m|$ of the pulsation mode. These relations are applicable to non-zonal ($m \neq 0$) spheroidal and toroidal modes and are valid also for multi-mode pulsations including those cases where temperature effects dominate over radial velocity effects. The blue-to-red phase difference of the mode yields $l$, whilst the phase difference for the first harmonic leads to the value of $|m|$ (Telting \& Schrijvers \cite{TS}).

For the $\nu_1$ pulsation mode of HD\,93521, the phase $\phi_1(\lambda)$ is a monotonic function of wavelength (see Figs.\,\ref{phase6678april2005}, \ref{fourmean5876} and \ref{four5876}), i.e.\ there are no huge changes in the slope over the interval where the significant line profile variations are detected. Therefore, this mode is unlikely to be an `outlier' in the sense defined by Telting \& Schrijvers (\cite{TS}) and the relation between $l$ and the blue-to-red phase difference $\Delta\phi$ inferred by the latter authors should thus be applicable to these modes\footnote{Telting \& Schrijvers (\cite{TS}) caution however that their relation was established for models with moderate rotation and hence stars that are not significantly flattened by rotation.}. Applying these relations to the phase differences given in Table\,\ref{tabspec} yields $l$ values of $8 \pm 1$ for $\nu_1$ and $4 \pm 1$ for $\nu_2$. 

Howarth \& Reid (\cite{HR}) interpreted the $\nu_1$ modulation as the signature of sectoral mode non-radial pulsations with $l = -m \simeq 9$. They further inferred a horizontal to radial velocity variation amplitude of $k < 0.3$. Later on, Howarth et al.\ (\cite{HTC}) derived $l \simeq 10 \pm 1$ and $6 \pm 1$ for $\nu_1$ and $\nu_2$ respectively, with $m + l \leq 2$. Whilst our values of $l$ are in rough agreement with those of Howarth \& Reid (\cite{HR}) and Howarth et al.\ (\cite{HTC}), the lack of a significant power in the first harmonics prevents us from deriving the value of $|m|$ from the simple scaling relations for this frequency. We note that the low amplitudes (or actually the upper limits on the amplitude) of the first harmonics compared to those of the genuine frequencies, suggest indeed that $|m|$ is likely close to $l$ (Schrijvers et al.\,\cite{schrijvers}). However, we stress that in a rapidly rotating early-type star, such as HD\,93521, the combined effects of a concentration of the pulsation towards the equator and of a non-uniform temperature distribution due to gravity darkening leads to more complicated amplitudes as a function of wavelength (e.g.\ Townsend \cite{Townsend}). A detailed line profile modelling is therefore required to derive the values of $m$ and we defer this to a forthcoming paper.

A priori, the shape of the semiamplitude of the modes as a function of wavelength (see Figs.\,\ref{phase6678april2005}, \ref{fourmean5876} and \ref{four5876}) suggests that the modes have a rather modest ratio between the amplitudes of the horizontal and radial velocity variations ($k \leq 0.5$). However, Schrijvers \& Telting (\cite{ST}) showed that the typical double-peaked shape of the amplitude found for velocity-dominated line profile variability with high $k$ values vanishes when temperature effects become important. Therefore, since we cannot exclude that the pulsation modes in HD\,93521 might be affected by temperature effects (see above), we cannot make a secure statement about $k$.

\begin{figure}[htb!]
\resizebox{9.0cm}{!}{\includegraphics{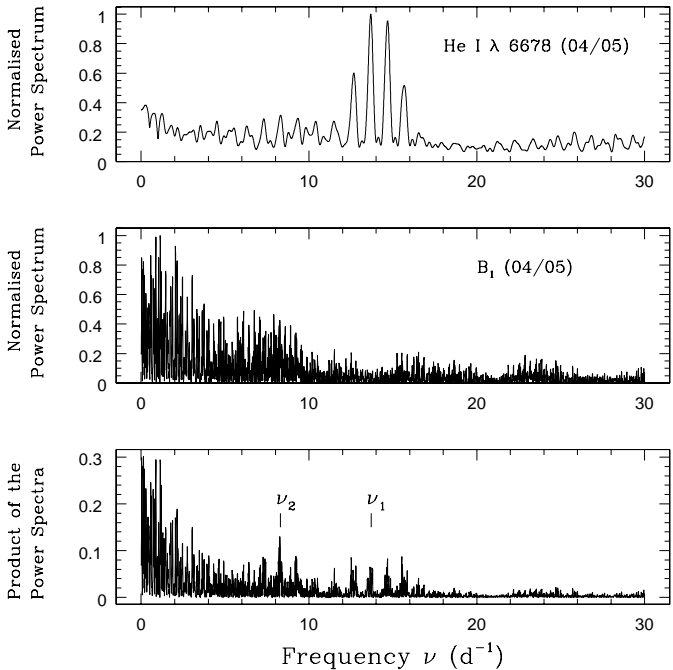}}
\caption{Top panel: normalised Fourier power spectrum of the He\,{\sc i} $\lambda$\,6678 line as observed in April 2005. Middle panel: normalised Fourier power spectrum of the $B_1$ photometric data obtained in April 2005. Bottom panel: product of the two normalised power spectra. \label{product}} 
\end{figure}

Our photometric data do not reveal significant variability at the $\nu_1$ and $\nu_2$ frequencies. However, because the $\nu_1$ mode is of high degree, the integrated flux variability due to these pulsations is indeed expected to be rather modest a priori and it could be that these variations just have too low an amplitude to be detected. To see whether or not this explanation is plausible, we have used the method outlined in Aerts et al.\ (\cite{deltaCeti}). First, we have normalized the periodogram of the April 2005 He\,{\sc i} $\lambda$\,6678 time series by dividing it by the power of the highest peak. The same procedure was then applied to the periodograms (for each filter) of the photometric data, also taken in April 2005. The normalized periodograms of the spectroscopic and photometric data were finally multiplied two by two with the rationale that frequencies that are present in both data sets should be dominant in the product of the periodograms (see Aerts et al.\ \cite{deltaCeti}). In this way, we find that the $\nu_2$ frequency is clearly seen in all the products, whilst it had a significantly lower amplitude than $\nu_1$ in the spectroscopic periodogram (see e.g.\ the case of the $B_1$ filter in Fig.\,\ref{product}). The $\nu_1$ frequency is seen with a strength slightly larger than the $\nu_2$ frequency only in the product of the spectroscopic and photometric periodograms of the $U$ and $G$ filters. The fact that $\nu_2$ appears more prominently in these products than $\nu_1$ would be consistent with our conclusion that the former mode has a lower $l$ value than the latter. Whilst this test is not a proof for the presence of the spectroscopic frequencies in the photometric variations, it nevertheless suggests that obtaining an extensive photometric time series with a lower noise level (e.g.\ using a space-borne observatory) would definitely be worth the effort.

We have further found several other possible periodicities that could be present in the photometric data. If real, these modulations might correspond to either radial or lower degree non-radial pulsations. Such modes are difficult to detect in rotationally broadened line profiles, but might well produce an observable signature in the photometric data. Walker et al.\ (\cite{Walker}) reported on {\it MOST} photometry of $\zeta$\,Oph: they found that the light curve is dominated by a 4.633\,hr period with a semiamplitude of 7.3\,mmag, whilst the other modes have semiamplitudes below 2.25\,mmag. These authors suggested that the light variations of $\zeta$\,Oph mainly result from radial pulsations. In the case of HD\,93521, the vast majority of the frequencies detected in the periodogram of the time series have semiamplitudes that are significantly lower than what was found in the case of $\zeta$\,Oph. Nevertheless, HD\,93521 would be a very interesting target for an intensive photometric monitoring from space. 
\subsection{The need for an alternative interpretation?}
Whilst multi-frequency non-radial pulsations offer an attractive interpretation of the line profile variability observed in the spectrum of HD\,93521, we must nevertheless ask the question whether there could be alternative explanations. We stress that the stability of the frequencies over many years (when they are detected) likely implies that the profile variations are produced by one or several stable clocks such as pulsations (considered above), rotation, orbital motion... Generally speaking, the distinction between these different mechanisms is a non-trivial issue (see e.g.\ the discussion in Uytterhoeven et al.\,\cite{Uytterhoeven}). For instance, in the case of  $\zeta$\,Oph, Harmanec (\cite{Harmanec}) proposed a different scenario where the moving bumps in the absorption lines actually stem from rotating inhomogeneities of the circumstellar material rather than from non-radial pulsations. 

In our case, there are also a number of reasons to consider alternative scenarios. Indeed, it is clear that currently a theoretical model to predict the line profile variations produced by non-radial pulsations in a rapidly rotating massive star such as HD\,93521 is still lacking. Therefore, the interpretation of these features within the framework of the available models requires an extrapolation that might be difficult to justify a priori. 

Another feature that is certainly puzzling is the fact that the line profile variability is significantly detected only in lines that are potentially affected by residual emission possibly associated with a circumstellar disk (or flattened wind). Whilst the lack of a TVS signal in the purely photospheric O\,{\sc iii} $\lambda$\,5592 line (see Fig.\,\ref{average}) might be interpreted as this line forming near the hotter poles of the star where the pulsational amplitude is lower (as for He\,{\sc ii} lines), the same explanation cannot hold for the photospheric features that occur at temperatures and gravities that are typical of those of the equatorial region (see the C\,{\sc ii}, N\,{\sc ii} and Si\,{\sc iii} lines discussed in Sect.\,\ref{spec}). The latter features produce at best a marginal signal in the TVS (see Fig.\,\ref{average}). However, regarding the possibility that the variability stems from rotating features in the circumstellar material, we first note that the frequencies $\nu_1$ and $\nu_2$ are not detected in the emission wings of the lines analysed in this work. If the line profile variations were coming from the flattened wind, one would also expect them to affect the emission wings. We further note that all the lines where we have detected line profile variability are quite strong. Actually, the rather low amplitude of the profile variations might render them undetectable in the weaker equatorial lines. 

In relation to this, we note that a recent study of HD\,60848 revealed evidence for the existence of rather short (3.51 and 3.74\,hr) periods in the radial velocities derived from the emission lines of this O9.5\,IVe star (Boyajian et al.\,\cite{Boyajian}). These authors argued that these features might result from {\it changes in the disk density or illumination caused by non-radial pulsations in the underlying star}. Therefore, it seems that a disk origin for short period variations cannot be ruled out a priori, although it would also reflect the existence of pulsations in this specific example. 

Since HD\,93521 was considered as a possible runaway object (Gies \cite{Gies}, but see also the discussion in Sect.\,\ref{intro}), it could host a compact companion which could then trigger a periodic structure in the inner part of the disk that is seen projected against the photosphere. However, the presence of a compact companion should make HD\,93521 a bright X-ray source, at least episodically when the compact object crosses the equatorial wind (similar to Be-type high-mass X-ray binaries). We thus checked the {\it ROSAT} All Sky Survey images as well as several other X-ray catalogues. To the best of our current knowledge, there is no indication that HD\,93521 is, or has ever been, a bright or even moderate X-ray emitter. Another point is that the detected frequencies $\nu_1$ and $\nu_2$ are much too high to correspond to the orbital period ($\geq 13.5$\,hr and likely of order several days) of such a putative companion. A compact companion scenario appears therefore rather unlikely. 

Another point concerns the fact that both frequencies are too high to correspond to the rotational period of the star. In fact, whilst many of the physical parameters of HD\,93521 remain unknown or poorly determined, we can obtain a rough estimate of the rotational frequency by considering typical parameters of a late O-type star. Let us assume that HD\,93521 is an O8.5 star\footnote{The observational classification as an O9.5 star is likely biased towards later spectral types as a result of gravity darkening due to the high rotational velocity and the nearly equator-on orientation of the star.} with a polar radius of 8\,R$_{\odot}$ and a mass of 20\,M$_{\odot}$. If the star is actually seen equator-on, the resulting rotational frequency would be 0.81\,d$^{-1}$. In general, we can say that for any reasonable assumption on the stellar parameters, we find that $\nu_{\rm rot} \leq 1$\,d$^{-1}$. Now, if the line profile variations actually stem from a regular pattern of moving spokes in the circumstellar disk, $\nu_1$ and $\nu_2$ would not necessarily have to correspond to $\nu_{\rm rot}$, but could rather be some harmonics of the latter (see Uytterhoeven et al.\,\cite{Uytterhoeven}). We would thus have to look for a ``super'' frequency such that $\nu_1 = n_1\,\nu_{\rm sup}$ and $\nu_2 = n_2\,\nu_{\rm sup}$, where $n_1$ and $n_2$ would be integer numbers. Within the uncertainties of our period determinations, a candidate for such a frequency would be $\nu_{\rm sup} = 2.75$\,d$^{-1}$ (with $n_1 = 5$ and $n_2 = 3$). Again, the latter frequency is much too high to correspond to the rotational frequency. The super-frequency could be in agreement with the likely value of the rotational frequency if $n_1$ and $n_2$ were muliplied by 3. However this would imply a large number (15 and 9) of co-rotating structures around the star which seems rather difficult to explain.
 
In summary, we conclude that, all the alternative scenarios envisaged here fail in explaining the modulations at the $\nu_1$ and $\nu_2$ frequencies. Hence, despite some difficulties, the multi-period non-radial pulsations model remains currently the most plausible explanation for the line profile variations seen in HD\,93521.

\acknowledgement{We thank the referee, Dr.\ D.\ Gies for his very helpful report. The Li\`ege group acknowledges financial support from the FRS-FNRS (Belgium), as well as through the XMM and INTEGRAL PRODEX contract (Belspo). The travels to OHP were supported by the `Communaut\'e Fran\c caise' (Belgium). The Mercator telescope and its operations are funded by the Catholic
University of Leuven, the Flemish community and the Fund for Scientific Research of Flanders (FWO). The Mercator observations were performed by the Leuven
team in the framework of the FWO project G.0178.02. PE acknowledges support through CONACyT grant 67041.}

\end{document}